\documentclass[usenatbib,useAMS,twocolumn]{mn2e}

\pdfoutput=1
\usepackage{subfigure}
\usepackage{graphicx}
\usepackage[latin1]{inputenc}
\usepackage{amssymb}
\usepackage{natbib}
\usepackage{color}

\bibpunct{(}{)}{;}{a}{}{,}
\newcommand{\apj}{ ApJ}
\newcommand{\aj}{ AJ}
\newcommand{\apjl}{ ApJL }
\newcommand{\mnras}{ M.N.R.A.S.}
\newcommand{\aap}{ Astron. Astrophys.}

\newcommand{\araa}{ ARAA}
\newcommand{\nat}{ Nature}
\newcommand{\physrep}{Phys. Rep.}
\newcommand{\qjras}{Q.J.R.A.S.}

\newcommand{\degree}{\ensuremath{^\circ}}
\def\ltsima{$\; \buildrel < \over \sim \;$}
\def\lta{\lower.5ex\hbox{\ltsima}}
\def\gtsima{$\; \buildrel > \over \sim \;$}
\def\simgt{\lower.5ex\hbox{\gtsima}}
\def \arcsec{$^{\prime\prime}$}

\title[ULIRGs, CO, and Universal Star Formation]{High-resolution CO and radio imaging of $z\sim2$ ULIRGs: extended CO structures, and implications for the universal star formation law }
\author[M. S. Bothwell et al. ]
{M. S. Bothwell$^{1}$\thanks{E-mail:
bothwell@ast.cam.ac.uk},
S. C. Chapman$^{1}$, 
L. Tacconi$^{2}$, 
Ian Smail$^{3},$
R. J. Ivison$^{4,5},$\newauthor
C. M. Casey$^{1},$
F. Bertoldi$^{6},$
R. Beswick$^{7},$
A. Biggs$^{8},$
A. W. Blain$^{9},$
P. Cox$^{10},$\newauthor
R. Genzel$^{2}$, 
T. R. Greve$^{11},$ 
R. Kennicutt$^{1},$ 
T. Muxlow$^{7},$
R. Neri$^{10},$ 
A. Omont$^{12}$\\
$^{1}$ Institute of Astronomy, University of Cambridge, Cambridge, CB3 0HA\\
$^{2}$ Max-Planck Institut f\"ur Extraterrestrische Physik, Giessenbachstrasse 1, D-85741 Garching, Germany\\
$^{3}$ Institute for Computational Cosmology, Durham University, Durham, DH1 3LE, UK\\
$^{4}$ UK Astronomy Technology Centre, Royal Observatory, Blackford Hill Edinburgh EH9 3HJ, UK\\
$^{5}$ Institute for Astronomy, University of Edinburgh, Blackford Hill, Edinburgh, EH9 3HJ, UK\\
$^{6}$ Argenlander Institute for Astronomy, University of Bonn, Auf dem H\"ugel 71, 53121 Bonn, Germany \\
$^{7}$ Jodrell Bank Centre for Astrophysics, School of Physics and Astronomy, University of Manchester, Oxford Road, Manchester, M13 9PL, UK\\
$^{8}$ European Southern Observatory, Garching D-85748, Germany\\
$^{9}$ Department of Astronomy, California Institute of Technology, 1200 E California Blvd, Pasadena, CA, 91125, U.S.A.\\
$^{10}$ Institut de Radio Astronomie Millimetriq\'ue, St. Martin d'Heres, France\\
$^{11}$ MPIA, K\"onigstuhl 17, D-69117 Heidelberg, Germany\\
$^{12}$ Institut d'Astrophysique de Paris, Universitte Pierre et Marie Curie and CNRS 98 bis boulevard Arago, 75014, Paris, France\\
}
\begin{document}
\date{Accepted 2010 February 4. Received 2010 February 3; in original form 2009 December 3}

\pagerange{\pageref{firstpage}--\pageref{lastpage}} \pubyear{2009}

\maketitle
\begin{abstract}
We present high spatial resolution (0.4\arcsec,  $\simeq 3.5$ kpc) PdBI interferometric data on three ultra-luminous infrared galaxies (ULIRGs) at $z\sim2$: two submillimetre galaxies and one submillimetre faint star forming radio galaxy. The three galaxies have been robustly detected in CO rotational transitions, either $^{12}$CO(J=4$\rightarrow$3) or $^{12}$CO(J=3$\rightarrow$2), allowing their sizes and gas masses to be accurately constrained. These are the highest spatial resolution observations observed to date (by a factor of $\sim2$) for intermediate-excitation CO emission in $z\sim2$ ULIRGs. The galaxies appear extended over several resolution elements, having a mean radius of 3.7 kpc. High-resolution (0.3\arcsec) combined MERLIN-VLA observations of their radio continua allow an analysis of the star formation behaviour of these galaxies, on comparable spatial scales to that of the CO observations. This `matched beam' approach sheds light on the spatial distribution of both molecular gas and star formation, and we can therefore calculate accurate star formation rates and gas surface densities: this allows us to place the three systems in the context of a Kennicutt-Schmidt (KS)-style star formation law. We find a difference in size between the CO and radio emission regions, and as such we suggest that using the spatial extent of the CO emission region to estimate the surface density of star formation may lead to error. This size difference also causes the star formation efficiencies within systems to vary by up to a factor of 5. We also find, with our new accurate sizes, that SMGs lie significantly above the KS relation, indicating that stars are formed more efficiently in these extreme systems than in other high-$z$ star forming galaxies.

\end{abstract}


\section{Introduction}
Recent evidence has suggested that the star formation activity of the universe peaks at redshifts between 1 and 3, and as such an understanding of the systems involved in forming stars at these cosmological redshifts is of great importance. Of particular interest are the luminous submillimetre galaxies, or SMGs \citep{2002MNRAS.331..495S}, which peak strongly at $z\sim2$ \citep{2005ApJ...622..772C}. First detected with the Submillimeter Common-User Bolometer Array (SCUBA, \citealt{1999MNRAS.303..659H}), these sub-mm selected (S$_{850 \mu \mathrm{m}} \simgt 3$ mJy) systems have prodigious star formation rates an order of magnitude greater than optically-selected galaxies at comparable redshifts ($\simgt$1000 M$_{\sun}$ yr$^{-1}$), large reservoirs of molecular gas ($\sim 10^{10} \,\mathrm{M}_{\sun}$; \citealt{2005MNRAS.359.1165G}), and are thought to contribute a substantial fraction of the total cosmic star formation rate density at their peak epoch ($\sim 50\%$ at $z\sim2$; \citealt{1999ApJ...512L..87B}, \citealt{2002PhR...369..111B}). Being sub-mm selected, the bolometric emission from these systems is dominated by cold dust ($\mathrm{T}_{\mathrm{d}} \sim 30$ K), and they are relatively faint at 24$\mu $m and in the X-ray, indicating that the contribution from an AGN component is generally negligible.

Observations of CO structures in SMGs have generally suggested relatively small spatial extents \citep{2008ApJ...680..246T}, typically having half light radii on the order 2 kpc. This, combined with the massive reservoir of gas in the interstellar medium (ISM), leads to a very high density (up to two orders of magnitude denser than their optically-selected counterparts - \citealt{2007ApJ...657..725N}) which triggers the intense star bursting behaviour that is characteristic of the SMG population. Many authors have concluded that the SMGs are therefore `scaled up' versions of ultraluminous infrared galaxies (ULIRGs, galaxies with L$_{\mathrm{FIR}} > 10^{12}\; \mathrm{L}_{\sun}$), in the local Universe, being a population of compact star forming galaxies driven by major mergers \citep{1996ARA&A..34..749S} with CO sizes comparable to the radii of the entire galaxy (\citealt{1999ApJ...524..732T}; \citealt{2007ApJ...659..283I}). In this model, the star formation is concentrated entirely in the dense circumnuclear region. Recent work has drawn upon evidence from features in the mid-IR (for example PAH emission in the study of \citealt{2009ApJ...699..667M}, and near-IR colours in the work of \citealt{2009ApJ...699.1610H}), using measures of the extinction to infer that the star forming region in SMGs is extended on scales of $> 2$ kpc; proportionally far larger than anything seen in local ULIRGs. It remains to be seen, therefore, whether conventional star formation prescriptions applicable to galaxies in the low redshift universe can account for the extreme star forming behaviour seen in SMGs.  

In this paper we report IRAM Plateau de Bure Interferometer (PdBI) observations of three high-redshift star forming sources. Combined MERLIN+VLA radio continuum observations are used to analyse the spatially resolved star formation behaviour (e.g. \citealt{2004ApJ...611..732C}; \citealt{2005MNRAS.358.1159M}; \citealt{2008MNRAS.385..893B}; \citealt{2009MNRAS.395.1249C}), in combination with the spatially resolved information on the molecular gas content provided by the PdBI, to attempt to analyse the connection between the gas content and star formation behaviour in these extreme high-redshift systems. The high-resolution of both the PdBI and MERLIN observations allow the sizes of the CO and star forming regions to be ascertained accurately and independently, allowing for accurate measurements of both the SFR surface density and the molecular gas surface density. As such, we can place our sources into the context of the Kennicutt-Schmidt law, which has been proposed as a possible `universal star formation law' \citep{2007ApJ...671..303B}.

In \S2, we outline the observations behind this work, and in \S2.3 we outline the techniques used to analyse our data. In \S3 we individually discuss our galaxies, in terms of their CO properties, star formation behaviour, and kinematical parameters. \S4 contains further discussion of merger vs. disc scenarios and star formation efficiencies for our galaxies, and attempts to place them in the context of a Kennicutt-Schmidt star formation law, and we conclude in \S5.

Throughout this work we assume a concordance $\Lambda$CDM cosmology with $h=0.71$, $\Omega_{\Lambda} = 0.72$, and $\Omega_{\mathrm{M}} = 0.28$.


\section[]{Observations and Analysis}
We aim to compare matched beam ($\sim0.4$\arcsec) observations of mid-excitation CO lines and radio continua in a selection of $z\sim2$ ULIRGs,  two in the Hubble Deep Field North and one in the Lockman hole, whose optical counterparts have been identified as HDF132 ($z =1.999$), HDF254 ($z =1.996$), and Lockman 38 ($z =1.523$) respectively. In this work they are referred to as such hereafter. This project was undertaken with a long-term goal of including similarly high resolution of the resolved sub-mm continuum in order to fully characterise the nature of star formation, gas content, and dust in these extreme systems. We have chosen three sources with good CO detections which have been well detected with MERLIN radio observations.  In particular, the radio properties of the three sources were chosen to be representative of the SMG radio population as a whole (see \citealt{2004ApJ...611..732C}), which has $\sim1/3$ being extended in the radio (HDF254), and $\sim2/3$ being radio compact (HDF132 and Lockman 38), so the large CO structures discussed below do not result from a selection bias. 

The three sources also span the complete range of $z\sim2$ ULIRG spectral energy distribution (SED) types - Lockman 38 suggests a unusually cold SED, with a dust temperature T$_{\mathrm{d}} \sim 20$K. HDF254 suggests a hot SED, with a T$_{\mathrm{d}}$ of $65 \pm 14$K \citep{2009arXiv0910.5756C}: it is strong in the radio, weak in the sub-mm, and has significant 70$\mu$m emission - it is, in fact, one of the few $z>1$ ULIRGs ($<5$\%) detected by Spitzer at 70$\mu$m. HDF132 is intermediate between these two sources, with an estimated T$_{\mathrm{d}}$ of 40K \citep{2005ApJ...622..772C}.

\subsection{PdBI CO observations}

We observed CO rotation lines in the sub-mm faint star forming radio galaxy (SFRG) RGJ123711.34+621331.0 (= HDF254), the sub-mm galaxy SMMJ123618.33+621550.5 (= HDF132), and the SMG SMMJ105307+572431.4 (= Lockman 38) during February and March 2009, using between 4 and 6 of the antennas (of a total of 6). All sources were observed with the `long baseline' A configuration.

HDF132 was observed in $^{12}$CO(J=4$\rightarrow$3) on 2009 February 15 and February 19, with on-source integration times of 8.9 and 2.9 hours respectively. Four of the six antennas were used. The observed frequency of the  $^{12}$CO(J=4$\rightarrow$3) line was 154.004 GHz, putting the CO-derived redshift at $z = 1.999 \pm 0.001$. HDF254 was observed in $^{12}$CO(J=4$\rightarrow$3)  on 2009 February 16 with a total integration time of 6.2 hours, using five antennas. $^{12}$CO(J=4$\rightarrow$3) was observed at 154.122 GHz, setting the source at $z = 1.996 \pm 0.001$. Lockman 38 was observed with all six antennas on 2009 February 20 and 2009 March 12, with on-source integration times of 5.6 hours and 3.2 hours respectively. $^{12}$CO(J=3$\rightarrow$2) was observed at 154.122 GHz, setting the source at $z = 1.523 \pm 0.002$. The synthesised beam size at 154 GHz is $\sim$ 0.4\arcsec (see panel insets), $\simeq$3.4 kpc at the redshifts of our sources.

The SMG SMMJ123711.98+621325.7 (= HDF255) was also within the field, being just 7\arcsec\ south-east of HDF254. It was, however, undetected with our long baseline configuration observations, which can be attributed to a combination of primary beam attenuation effects, and the CO source being sufficiently extended that all the flux is resolved out, consistent with a large radio size ($\sim 3.2$ kpc)  from MERLIN as reported by \cite{2004ApJ...611..732C}. We include HDF255 in our compilation of SMGs from the literature, however, and use the fact that it was undetected in our high resolution observations to place a lower limit on the source diameter of 1\arcsec\ (hence, all surface densities derived for HDF255 will strictly be upper limits).

All sources were observed in good meteorological conditions, and typical system temperatures were 80-250K. Antenna efficiencies were typically 25-35 Jy K$^{-1}$.  Bright quasars (primarily 3C84 and 3C273) were used for flux calibration. All data were reduced using the \textsc{clic} and \textsc{mapping} routines as part of the IRAM \textsc{gildas} package, and the resultant data cubes were rebinned to a frequency resolution of 20 MHz (a velocity spacing of 38 km s$^{-1}$), before analysis with our own IDL routines.

\subsection{MERLIN 1.4 GHz continuum observations}
High resolution observations of the galaxies' radio continua were obtained using the Multi-Element Radio Linked Interferometer Network (MERLIN; \citealt{1986QJRAS..27..413T}), combined with the VLA. Radio data for the HDF sources were obtained by \cite{newvlafluxes}, and data for Lockman 38 were obtained by \cite{2009MNRAS.397..281I}.

The FWHM synthesised beam sizes for the maps are (semi-major axis vs. semi-minor axis in arcsec) 0.204 x 0.193, 0.205 x 0.192, and 0.5 x 0.5 for HDF132, HDF254, and Lockman 38 respectively. These beam sizes are all comparable in resolution to the long-baseline configuration beam size of the PdBI. The high spatial resolution of these observations allows for the star formation regions to be mapped in detail (rather than returning a single integrated value for the SFR), which is useful both for distinguishing point source AGN emission from diffuse star formation, and for comparing the extent and shape of the star forming region with the CO morphology (see \citealt{2008MNRAS.385..893B}). All three of our galaxies were well detected and fully resolved. 

\subsection{Analysis}
\label{gannon}

\begin{figure*}
 \centerline{\includegraphics[scale=0.5]{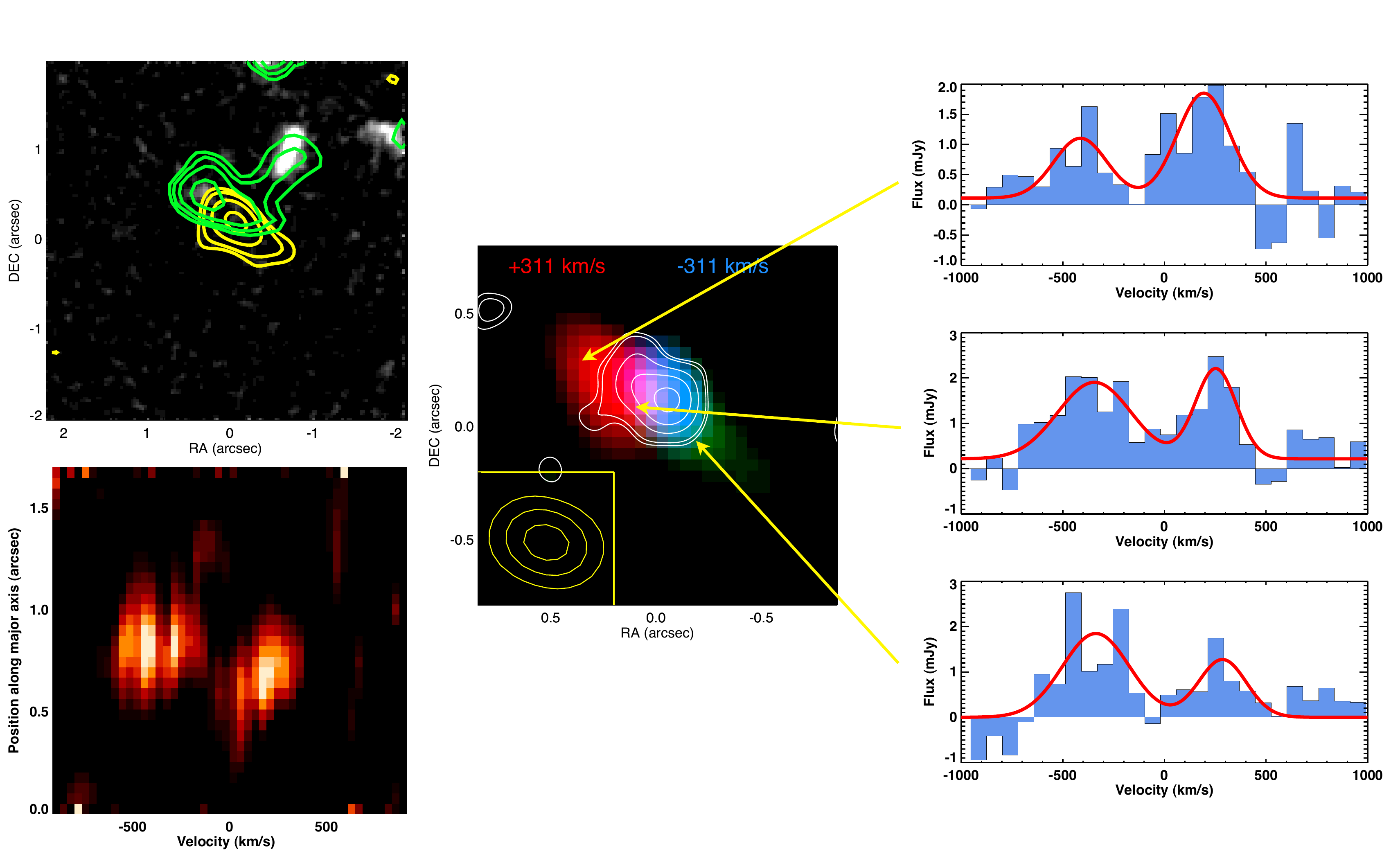}}
\caption{HDF132. \textit{Top Left:} Stacked ACS image with overlaid CO emission contours (yellow) and NICMOS F160W contours (green).  The yellow contours shown in the left hand panel delineate the 97, 98, 99, 99.5 and 99.9th percentiles of the velocity integrated CO data (the resultant levels in mJy/beam are 0.43, 0.50, 0.74, 1.18, and 1.90). \textit{Bottom Left}: Position-velocity diagram taken along the major axis of the galaxy (PA = 228\degree; defined by the position angle of the best fitting ellipse to the CO map), with size and velocity scales shown inset. \textit{Centre}: Velocity-coded CO emission, with overlaid MERLIN 1.4GHz radio contours (white; levels in $\mu$Jy/beam are 10, 12, 34, 40, and 65). The inset velocity shift legend represents the \textit{centre} of either the `red' or the `blue' bin, respective to the centre of the `green' central bin. The CO has been plotted starting at a significance level of 2$\sigma$ (in the integrated map); all points with detections $<2\sigma$ have been left black. The radio beam has semi-major and semi-minor axes of 0.204\arcsec\ $\times$ 0.193\arcsec\ respectively. The outer contour of the inset ellipse (bottom left) corresponds to the FWHM of the CO beam, which is $\sim 0.4$\arcsec. Velocities are coded relative to the systemic velocity, and are shown inset. \textit{Right:} 1D spectra, binned to 40 MHz, taken at the positions indicated. The spectra have been fitted with double-peaked Gaussian profiles.} 
\label{fig:sab1}
\end{figure*}

\begin{figure*}
 \centerline{\includegraphics[scale=0.5]{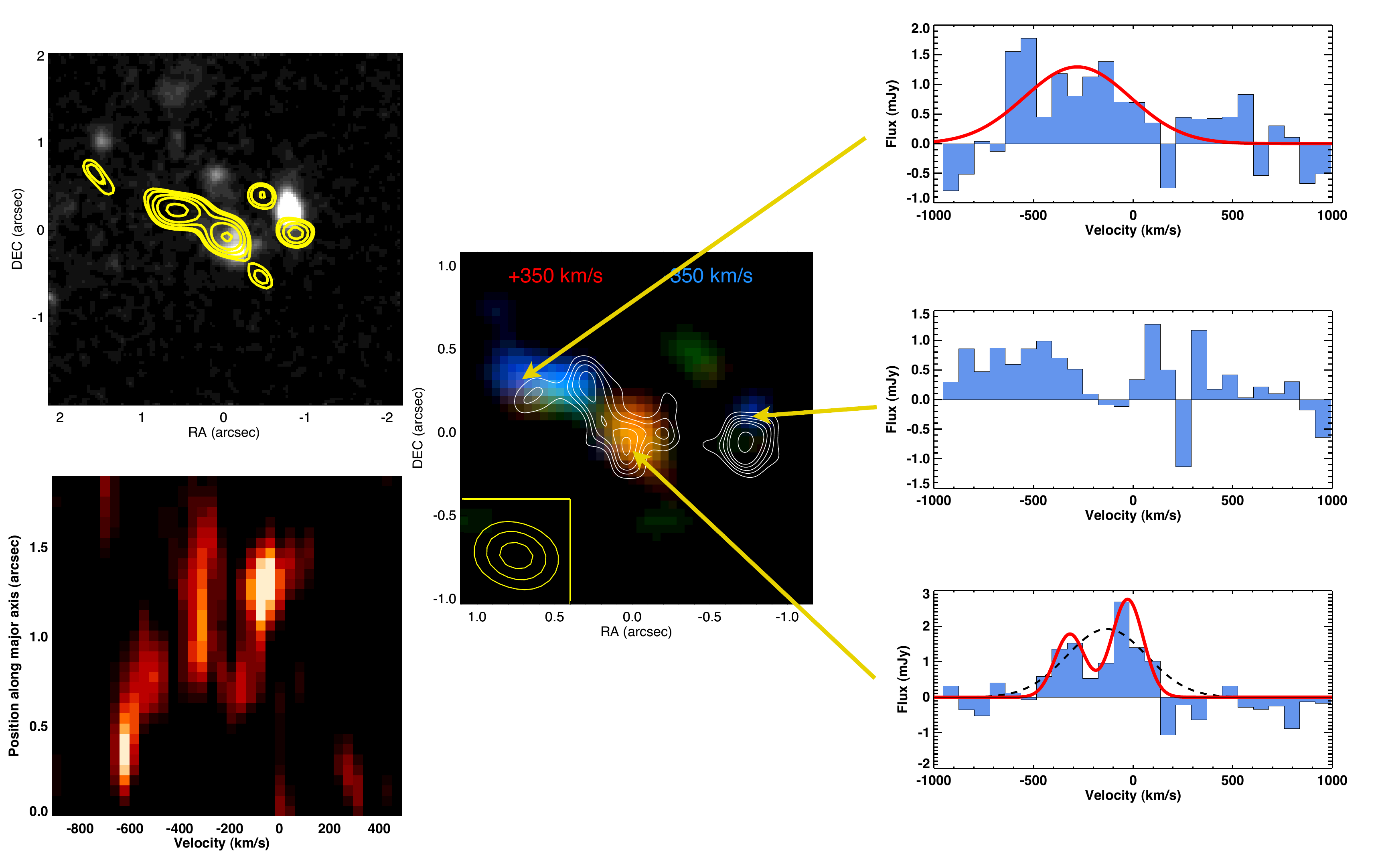}}
\caption{HDF254. \textit{Top Left:} Stacked ACS image with overlaid CO emission contours (yellow). The yellow contours shown in the left hand panel delineate the 97, 98, 99, 99.5 and 99.9th percentiles of the velocity integrated CO data (the resultant levels in mJy/beam are 0.40, 0.46, 0.56, 0.67, and 0.85). \textit{Bottom Left}: Position-velocity diagram taken along the major axis of the source (PA = 237\degree;  defined by the position angle of the best fitting ellipse to the CO map), with size and velocity scales shown inset. \textit{Centre}: Velocity-coded CO emission, with overlaid MERLIN 1.4Ghz radio contours (white; levels in $\mu$Jy/beam are 16, 18, 21, 24, and 33). The inset velocity shift legend represents the \textit{centre} of either the `red' or the `blue' bin, respective to the centre of the `green' central bin. The CO has been plotted starting at a significance level of 2$\sigma$ (in the integrated map); all points with detections $<2\sigma$ have been left black. The radio beam has semi-major and semi-minor axes of 0.205\arcsec\ $\times$ 0.192\arcsec\ respectively. The outer contour of the inset ellipse (bottom left) corresponds to the FWHM of the CO beam, which is $\sim 0.4$\arcsec. Velocities are coded relative to the systemic velocity, and are shown inset. \textit{Right:} 1D spectra, binned to 40 MHz, taken at the positions indicated. The spectra have been fitted with either single- or double-peaked Gaussian profiles (the fit with the lowest $\chi^2$ was used in each case).}
\label{fig:sdb1}
\end{figure*}

\begin{figure*}
 \centerline{\includegraphics[scale=0.5]{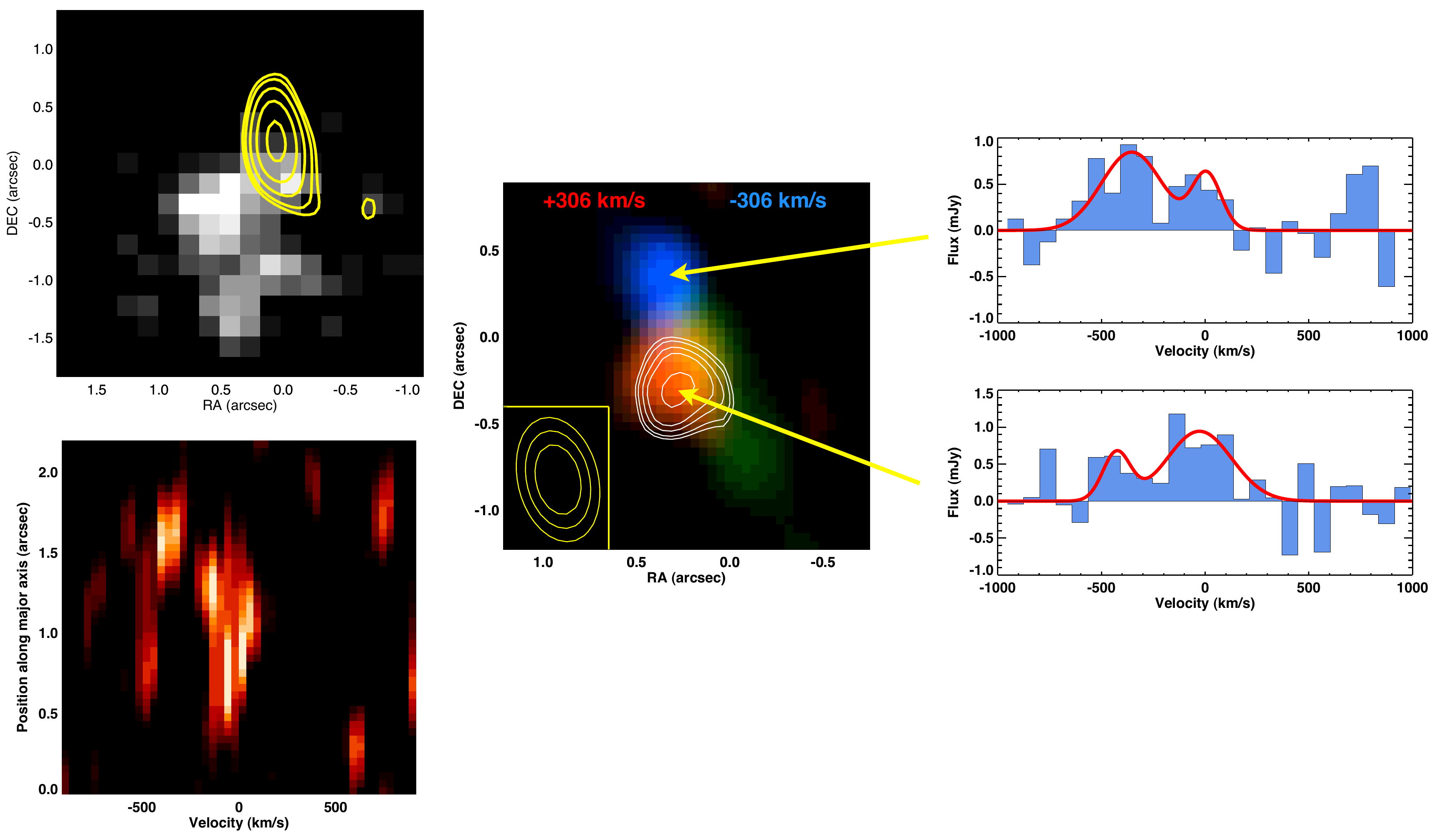}}
\caption{Lockman 38. \textit{Top Left:} CO emission contours (yellow), over CFHT-CFH12K $I$-band (greyscale).  The yellow contours shown in the left hand panel delineate the 97, 98, 99, 99.5 and 99.9th percentiles of the velocity integrated CO data (the resultant levels in mJy/beam are 0.23, 0.26, 0.32, 0.40, and 0.49).  \textit{Bottom Left}: Position-velocity diagram taken along the major axis of the source (PA = 193\degree; defined by the position angle of the best fitting ellipse to the CO map), with size and velocity scales shown inset.\textit{Centre}: Velocity-coded CO emission,  with overlaid MERLIN 1.4Ghz radio contours (white; levels in $\mu$Jy/beam are 7, 8, 9, 10, and 13). The inset velocity shift legend represents the \textit{centre} of either the `red' or the `blue' bin, respective to the centre of the `green' central bin. The CO has been plotted starting at a significance level of 2$\sigma$ (in the integrated map); all points with detections $<2\sigma$ have been left black. The radio beam has a radius of $0.5$\arcsec. The outer contour of the inset ellipse (bottom left) corresponds to the FWHM of the CO beam, which is $\sim 0.5$\arcsec. Velocities are coded relative to the systemic velocity, and are shown inset. \textit{Right:} 1D spectra, binned to 40 MHz, taken at the positions indicated. The spectra have been fitted with double-peaked Gaussian profiles. }
\label{fig:sfb1}
\end{figure*}

Figs \ref{fig:sab1}, \ref{fig:sdb1}, and \ref{fig:sfb1} show the CO observations and spectra, combined with optical images and MERLIN+VLA radio continuum observations (as detailed in the captions). Tables \ref{tab1} and \ref{tab2} contain the observational results and other derived parameters. All lines were successfully detected at a very high level of significance ($> 6 \sigma$). 

The two HDF sources (132 and 254) have optical images comprised of $b, v, i$ \& $z$ band images from the ACS, stacked with equal weighting. In the case of HDF132, NICMOS 1.6 micron photometry (using the F160W filter) has been added for clarification. There is no ACS image for Lockman 38; instead, we use an $I$ band image, taken with the CFH12K instrument on the Canada-France-Hawaii Telescope (CFHT). Being ground-based, this is necessarily a lower-resolution image than the \textit{HST}-ACS photometry, with a resolution of $\sim 0.7$\arcsec.

The middle panels of Figs \ref{fig:sab1}, \ref{fig:sdb1}, and \ref{fig:sfb1} also show the CO data colour coded by velocity. These were created as follows: each data cube has a channel range across which the signal to noise of the source is maximised. For each source, this channel range was then split into three equal bins, and the CO flux for each point on the map was integrated for each bin separately. This results in three separate maps of the source, corresponding to the three bins in channel space (or three bins in velocity space). These three maps were then colour-coded appropriately and stacked. 

The star formation rate given in Table \ref{tab2} has been derived using the empirically calibrated FIR-radio correlation, taken from \cite{2001ApJ...554..803Y}\footnote{The radio flux densities used here were obtained from the VLA-only data which recover the total flux density of each source, and are more sensitive to diffuse, extended regions of emission which may be resolved out by the longer baseline MERLIN data.}, 
\begin{equation}
L_{FIR} = 4\pi D_L^2 \;(8.4 \times 10^{14})\; S_{1.4} \;(1+z)^{(\alpha - 1)},
\end{equation}
where $D_L$ is the luminosity distance in metres, $S_{1.4}$ is the radio flux density at 1.4 GHz (in W m$^{-2}$ Hz$^{-1}$), and $\alpha$ is the synchrotron slope used to K-correct the 1.4 GHz observations to the appropriate source redshift ($\alpha$ is taken here to be 0.8). The L$_{\mathrm{FIR}}$-SFR conversion was taken from \cite{1998ARA&A..36..189K}:
\begin{equation}
\mathrm{SFR} = 4.5 \times 10^{-44}\; L_{FIR} \;(\mathrm{ergs}\, \mathrm{s}^{-1}).
\end{equation}
It is important to note that several assumptions underlie the direct conversion between $\mathrm{L}_{\mathrm{FIR}}$ and a SFR; namely that all the UV radiation produced by young stars is absorbed by dust and re-radiated in the IR, and that the dust heating is brought about by a young stellar population ($\leq 30$ Myr). If the stellar population is more evolved, the true SFR can be lower than that derived using the equation above by up to a factor of 2.

CO flux densities were estimated by fitting a Gaussian profile to the spectrum. From this, we calculate the CO luminosities using the relation given by \cite{1997ApJ...478..144S};
\begin{equation}
\mathrm{L}^{\prime}_{\mathrm{CO}} = 3.25 \times 10^7 \;S_{\mathrm{CO}}\; \nu_{\mathrm{obs}}^{-2}\; (1+z)^{-3}\;D_{L}^2
\end{equation}
(where $S_{\mathrm{CO}}$  is the CO flux in Jy km s$^{-1}$, $\nu_{\mathrm{obs}}$ is the frequency of the observed CO line in GHz, and the luminosity distance $D_{{L}}$ is in Mpc), from which molecular gas masses were derived using a conversion factor appropriate to local ULIRGs of $X= 0.8\; \mathrm{M}_{\sun}\; \mathrm{K  \;  km\; s}^{-1} \mathrm{pc}^{2}$ (see \citealt{2008ApJ...680..246T} for discussion).

These molecular gas masses have been derived from CO luminosities resulting from moderate-excitation transitions (either 4$\rightarrow$3 or 3$\rightarrow$2); the extrapolation of an accurate gas mass, therefore, requires a derivation of a CO(1$\rightarrow$0) flux density. We use the brightness temperature conversions derived from SED modelling of SMGs by \cite{2007ASPC..375...25W}, which leads to the following line ratios (as reported by \citealt{2009arXiv0910.5756C}): S$_{\mathrm{CO3-2}}$/S$_{\mathrm{CO1-0}} = 7.0\pm0.5$; S$_{\mathrm{CO4-3}}$/S$_{\mathrm{CO1-0}} = 10.0\pm0.8$. Most ULIRGs are thermalised up to at least the CO(3$\rightarrow$2) transition, so using these mid-excitation lines (J$_{\mathrm{upper}} =$ 3 or 4) should not significantly reduce the extent of the CO emission region when compared to observations of CO(1$\rightarrow$0). Any size difference will be due to the fact that the higher order transitions trace the actively star forming gas, (which is of interest to us) whereas the lower transitions trace the cold gas either in the outer halo, or accreting onto the galaxy. 

Use of the long baseline `A' configuration will inevitably resolve out flux when compared to observations made with more compact configurations. For HDF132 and HDF254 we compare our gas masses derived from our observations to those derived using the PdBI `D' configuration, to get an estimate of how much flux has been lost, which is typically a factor of $\sim2$. For Lockman 38, which has not been observed in `D' configuration, we estimate a true gas mass based on the mean flux lost for the other two sources. We use the (more complete) `D' configuration-derived gas masses when calculating gas surface densities for all systems. 

\section{Results}

\begin{table*}
\centering
\caption{CO properties for HDF132, HDF254, and Lock38. Note that HDF132 and Lock38 have two entries, for values derived from each of their double peaks. *H$_2$ masses derived from compact-configuration data, to avoid resolving out flux.}
\begin{tabular}{cccccccccc}
\hline
\hline
Source  &RA&DEC& z$_{\mathrm{CO}}$ &  S$_{\mathrm{CO}}$ & L$^{\prime}$ $_{\mathrm{CO}}$ & L$^{\prime}$ $_{\mathrm{CO(1-0)}}$ &  M(H$_2$)*\\
 & [J2000] & [J2000]  &&[Jy km s$^{-1}$] & [$10^{10}$ K km s$^{-1}$ pc$^{2}$] & [$10^{10}$ K km s$^{-1}$ pc$^{2}$] &  [10$^9$] \\
 \hline
 \hline
HDF132  &12:36:18.337 & 62:15:50.45& 1.996  & $0.77 \pm 0.12$ & $0.94 \pm 0.14$& $0.59 \pm 0.11$ & $7.2 \pm 1.5$\\
&&&2.001&$0.50\pm0.08$&$0.65\pm0.09$& $0.40 \pm 0.07$&&\\
\hline
HDF254  &12:37:11.354&62:13:30.90& 1.996  &$0.63 \pm 0.09$ & $0.81 \pm 0.11$& $0.51 \pm 0.09$& $8.3 \pm 1.7$\\
\hline
Lockman 38 &10:53:07.100 & 57:24:31.87& 1.523  & $0.51 \pm  0.07$ &$0.68 \pm 0.49$&$0.53 \pm 0.36$ & $8.0 \pm 3.9$&\\
&&&1.519&$0.12\pm0.03$&$0.16\pm0.04$&$0.12 \pm 0.03$&&\\
\hline
\end{tabular}
\label{tab1}
\end{table*}

\begin{table*}
\centering
\caption{Other properties for HDF132, HDF254, and Lock38. Note that FIR luminosities and SFRs are derived from the 1.4 GHz radio flux densities. Note also that the SFR for HDF254 has been corrected for AGN contamination, as described in \S\ref{254}. The 1.4 GHz radio fluxes were obtained by Morrison et al. (prep). }
\begin{tabular}{cccccccccc}
\hline
\hline
Source  & S$_{850}$& S$_{1.4}$ & L(FIR) &  SFR(FIR) &$\log \Sigma_{\mathrm{SFR}}$  & CO size (radii) & 1.4GHz size (radii)& M$_{\mathrm{dyn}}$ \\
 & [mJy]&[$\mu$Jy]&[$10^{12}$ L$_{\sun}$ ]&[M$_{\sun}$ yr$^{-1}$]&[M$_{\sun}$ yr$^{-1}$ kpc$^{-2}$] &[a\arcsec $\times$ b\arcsec]&[a\arcsec $\times$ b\arcsec]  M$_{\sun}$] &[$10^{10} $M$_{\sun}$]\\
 \hline
 \hline
HDF132 & $7.3\pm1.1$&$172\pm8.4$& $8.6\pm0.6$  & $1500\pm100$ & 2.19 &0.39\arcsec $\times$ 0.24\arcsec &0.24\arcsec $\times$ 0.18\arcsec & $21.8\pm1.8$ \\
\hline
HDF254 &$2.5\pm1.2$& $126\pm8.7$& $6.5\pm0.8$  & $1050\pm90$  & 1.76 &0.52\arcsec $\times$ 0.16\arcsec &0.41\arcsec $\times$ 0.14\arcsec&$9.6\pm0.5$  \\
\hline
Lock38 &$6.5\pm1.9$& $56.2\pm22.4$& $1.5\pm0.6$  & $250\pm80$  & 1.13 &0.56\arcsec $\times$ 0.31\arcsec &0.29\arcsec $\times$ 0.25\arcsec &$6.2\pm1.1$  \\
\hline
\end{tabular}
\label{tab2}
\end{table*}

\begin{figure*}
 \centerline{\includegraphics[scale=0.5]{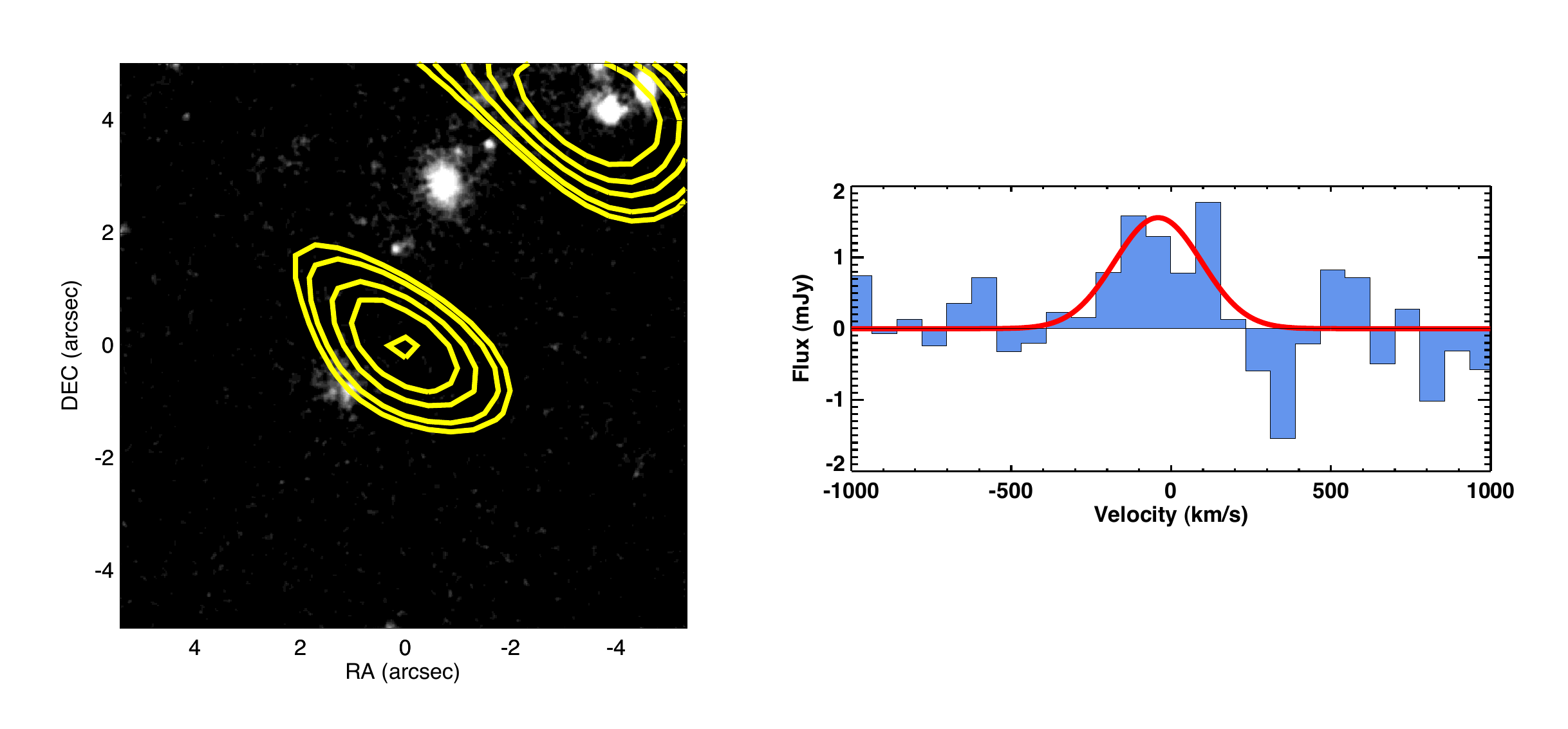}}
\caption{HDF255. Left: Stacked ACS image with overlaid CO emission contours, taken with the compact `D' configuration with PdBI (yellow). For reference, the source in the top right is HDF254. Right: CO spectra, also taken in compact `D' configuration, at the centre of the source. For reference, the strongly detected source to the NW of HDF255 (which is the source at the centre of the image) is HDF254.}
\label{fig:255}
\end{figure*}


\subsection{HDF132}
\label{132}
HDF132 is both the brightest of our three sources at millimetre wavelengths, and the most highly obscured (being virtually undetected in the optical), implying an extremely high column density of dust. However, it is detected in the IR (as evidenced by the NICMOS F160W imaging, which was obtained by \citealt{2008MNRAS.386..909C} as part of PID $\#$11082), and is sub-mm bright, with a SCUBA 850$\mu$m flux of $7.3\pm1.1$ mJy. The strong 1.4 GHz continuum luminosity implies a high rate of extinction-corrected star formation. This is calculated (from the radio continuum, using the methods described above) to be $1500\pm100$ M$_{\sun}$ yr $^{-1}$ (here and throughout, quoted errors on the SFR represent fractional errors in the measured radio flux density, and do not take into account other uncertainties, for example scatter in the L$_{\mathrm{FIR}}$-radio correlation).  The velocity-coded image in the centre of Fig. \ref{fig:sab1} shows what appears to be either a disc-type structure or a late stage merger.

Spectra taken at separate points along the source (see positional arrows in Fig. \ref{fig:sab1}) reveal the CO spectrum to be double peaked at all points, with the relative intensity of the peaks varying along the axis of the source. (The spectra have been more coarsely binned, to 40 MHz [$\sim 76$ km s$^{-1}$] for clarity). This could possibly be indicative of a `beam smearing' effect, whereby the size of the beam (see inset ellipse) causes both velocity components to be detected. However, the position in velocity space of each peak is approximately constant across the galaxy, with just their respective strengths varying. This, coupled with the relatively small beam size (0.4\arcsec\ - smaller than the source), suggests that the two peaks may be components of line of sight rotation.

The position-velocity diagram (Fig. \ref{fig:sab1}, bottom left) shows evidence for ordered gas motion, with two distinct clumps separated in velocity space; while they do overlap in physical space they are offset somewhat (resulting in the two peaks changing in intensity along the axis), this could be a signature of rotation (e.g. \citealt{2002A&A...388...50F}), but also may be indicative of a late stage merger.

\begin{figure}
\centerline{\includegraphics[scale=0.55]{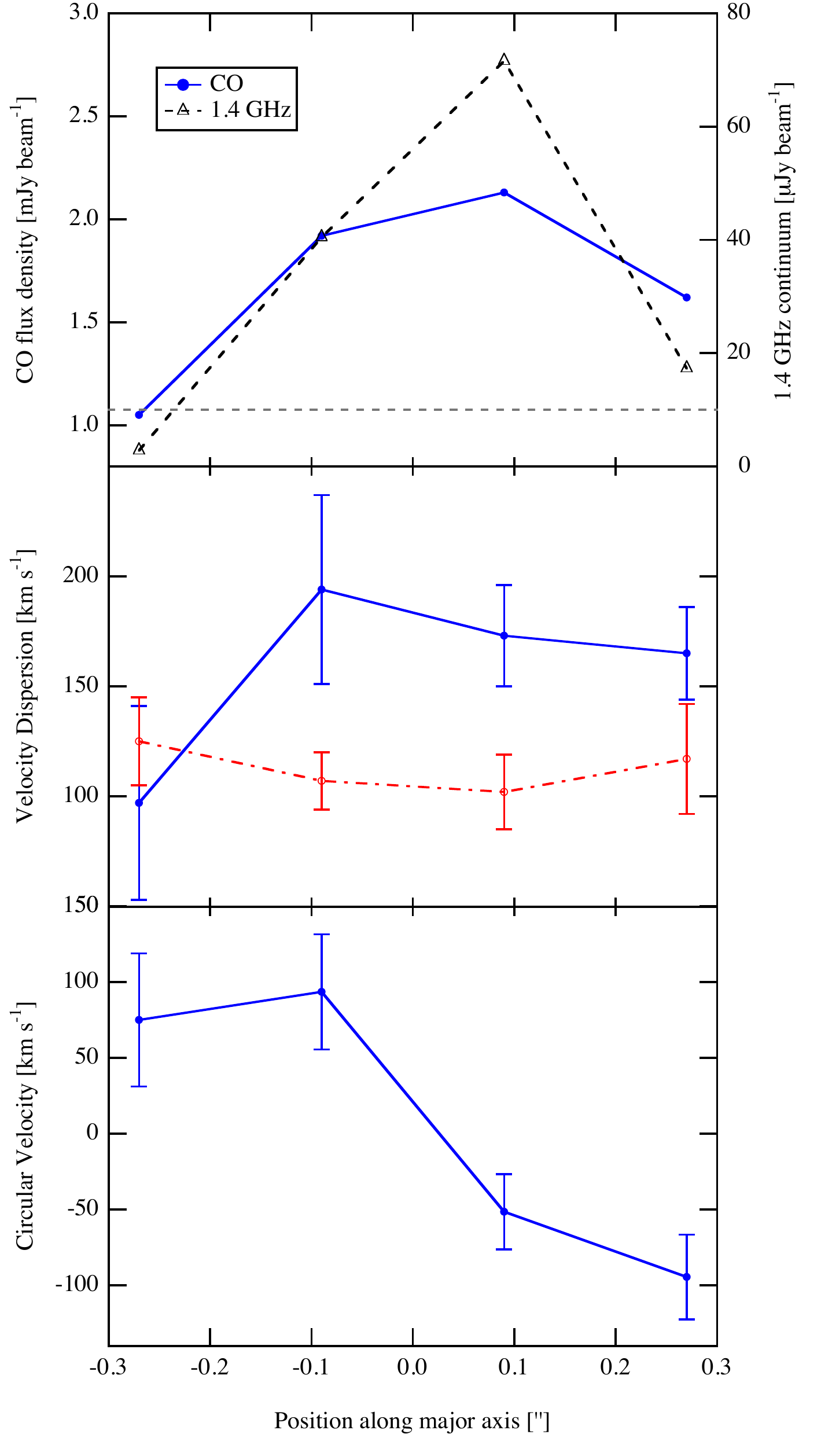}}
\caption{Kinematic analysis for HDF132 (SMMJ123618.33+621550.5). Spectra were taken at points along the major axis of the galaxy with $\sim$0.15\arcsec\ spacing ($\simeq$ 1/3 - 1/2 of the beam FWHM), and resultant kinematic parameters were derived. \textit{Bottom panel}: circular velocity; as the CO spectrum for HDF254 is double peaked, the velocity was obtained by taking the mean of the two peaks, with the weighting of each component taken as the height of the respective peak . \textit{Middle panel}: velocity dispersion, obtained via the width of the same Gaussian peak. The solid blue line shows the blue shifted peak, whereas the dashed red line shows the red shifted peak. \textit{Top panel}: CO surface brightness (blue solid line). For reference, the MERLIN-obtained 1.4GHz continuum flux density measured at the same points has been overlaid (black dashed line and right hand axis). The dashed horizontal line is drawn at 10$\mu$Jy/beam ($\sim 3$ times the image rms)..}
\label{fig:sab1_k}
\end{figure}

Fig. \ref{fig:sab1_k} shows the kinematics of HDF132 in more detail. Data were taken along the major axis of the source at $\sim 0.15$\arcsec\ intervals (or $\sim 1/3 - 1/2$ of the beam FWHM). The bottom panel shows the velocity at each point, obtained by fitting Gaussian curves to each peak of the source. To reach the final mean value for the velocity spread the positions of the two peaks were weighed by their respective strengths. It is likely that the value of the velocity shear obtained in this manner underestimates the true value, as beam smearing effects cause there to be significant emission from both components at most points. A better estimate for the true $v_{\mathrm{d}}$ would be $\Delta v / 2$, where $\Delta v$ is the separation of the two peaks in velocity space. For HDF132, $\Delta v / 2 = 288\pm 19$ km s $^{-1}$.

The middle panel of Fig. \ref{fig:sab1_k} shows the velocity dispersion at each point, taken to be the 1$\sigma$ value of the Gaussian fit to the peak. The `red' and `blue' peaks have been treated separately, and are coloured red and blue respectively. The top panel compares the CO and the 1.4 GHz flux density at each point. The small size of the source (D $<$ 1\arcsec) only allows for four points to be taken with enough S/N to allow robust spectra to be taken. This precludes the full elucidation of the velocity behaviour that would be able to characterise the source as either a rotating disc or a merger. However, using the velocity dispersion (Fig. \ref{fig:sab1_k}, middle panel) it is possible to estimate the lower limit on $v_d / \sigma$ (see \citealt{2008ApJ...687...59G}), a simple way of characterising the system as either `rotation-dominated' ($v_d / \sigma >2-3)$ or `dispersion-dominated' ($v_d / \sigma < 1$)\footnote{For reference, $z\sim0$ discs typically have ($v_d / \sigma$) $\sim10-20$ \citep{2006ApJ...638..797D}, whereas high-$z$ star forming discs have ($v_d / \sigma$) of order unity to a few (\citealt{2008ApJ...687...59G}; \citealt{2008Natur.455..775S}).}. The mean velocity dispersion is $113\pm19$ km s$^{-1}$  and $157\pm32$ km s$^{-1}$ for the red and blue components respectively, leading to a $v_d (\sin i)/ \sigma = 2.03\pm0.26$. This is, to first order, comparable to the value claimed for other high redshift star forming discs (e.g. \citealt{2006ApJ...645.1062F}; \citealt{2006Natur.442..786G}; \citealt{2007ApJ...669..929L}). 

Given the fact that it is not apparently dispersion-dominated, and that it is homogeneous, compact, and non-clumpy on scales $<0.5$\arcsec , we identify HDF132 as a disc-type structure, as opposed to an early-stage major merger (although of course a late-stage major-merger remnant is very difficult to distinguish from a true secular disc given the limitations of our observations).

The top panel of Fig. \ref{fig:sab1_k} shows a spatial breakdown of the CO and 1.4 GHz distributions. The two trace each other closely, both being centrally peaked and concentrated most strongly over $\sim 0.5$\arcsec, and reaching a maximum value at the same point. The radio continuum emission peaks more strongly, however, falling off more sharply than the CO emission either side of the central peak. Based on its CO morphology, we assign a source size (based on the CO emission half light radius\footnote{Note that \cite{1998ApJ...498..541K} used the \textit{full} measured extent of the emission to calibrate the KS law, whereas we use a `half light' radius as standard for interferometric observations. Any bias introduced by this difference should be negligible.}) of 0.39\arcsec\ by 0.24\arcsec, = 3.3 $\times$ 2.0 kpc at $z=1.996$. 

The long-baseline `A' configuration observations often resolve out CO flux when compared to a lower-resolution observation (such as one made in compact `D' configuration with the PdBI). To ensure that we deal with accurate gas surface densities, we take the source size from the high-resolution observations as described above, and the total gas mass from low-resolution (compact configuration) observations. The total gas mass for HDF132 is $7.2\times 10^9 \, \mathrm{M}_{\sun}$.  (from as-yet unpublished PdBI observations )\footnote{This suggests that the high-resolution observations have resolved out a factor of 1.8 in flux.}. This leads to a gas surface density (taking the total gas mass from a mean of the two component peaks) of $\Sigma_{\mathrm{H2}}$ = 540 M$_{\sun}$ pc$^{-2}$. To compare, the star-forming region (R$_{1/2}$ derived from the radio source extent) is 0.24\arcsec\ by 0.18\arcsec. The star formation surface density based on the FIR-inferred SFR is therefore $\Sigma_{\mathrm{SFR}}$ = 150 M$_{\sun}$ yr $^{-1}$  kpc$^{-2}$. 

Given that we have identified HDF132 as a disc-type structure, we can use a simple disc rotation model to calculate its dynamical mass: given its radius of 3.52 kpc and peak-to-peak rotation velocity of $288\pm 19$ km s $^{-1}$, we calculate a dynamical mass of M$_{\mathrm{dyn}}\sin^2 i = 5.5 \pm0.5 \times 10^{10}\;\mathrm{M}_{\sun}$ (it is important to note that this is assuming the gas is undergoing Keplerian motion). Adopting a fiducial inclination angle of 30\degree, we find the dynamical mass to be M$_{\mathrm{dyn}}= 2.2\pm0.2 \times 10^{11}\;\mathrm{M}_{\sun}$.

\begin{figure}
\centerline{\includegraphics[scale=0.55]{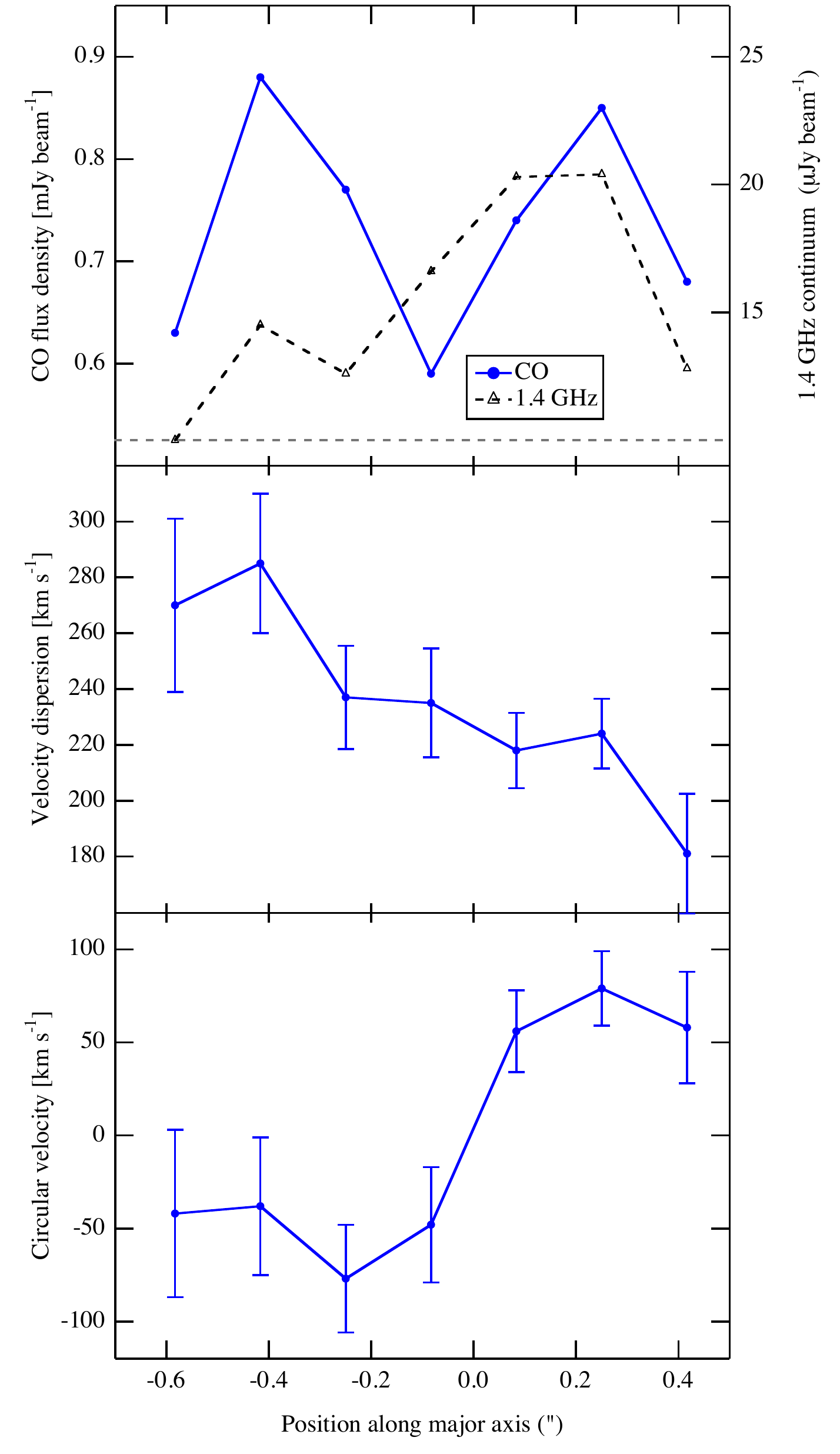}}
\caption{Kinematic analysis for HDF254. Spectra were taken at points along the major axis of the galaxy with $\sim$0.15\arcsec\ spacing ($\simeq$ 1/3 - 1/2 of the beam FWHM), and resultant kinematic parameters were derived. \textit{Bottom panel}: circular velocity, obtained using the centroid of the Gaussian profile fit to the CO emission line. \textit{Middle panel}: velocity dispersion, obtained via the width of the same Gaussian peak. \textit{Top panel}: CO surface brightness (blue solid line). For reference, the MERLIN-obtained 1.4GHz continuum flux density measured at the same points has been overlaid (black dashed line and right hand axis). The dashed horizontal line is drawn at 10$\mu$Jy/beam ($\sim 3$ times the image rms)..}
\label{fig:sdb1_k}
\end{figure}

\begin{figure}
\centerline{\includegraphics[scale=0.55]{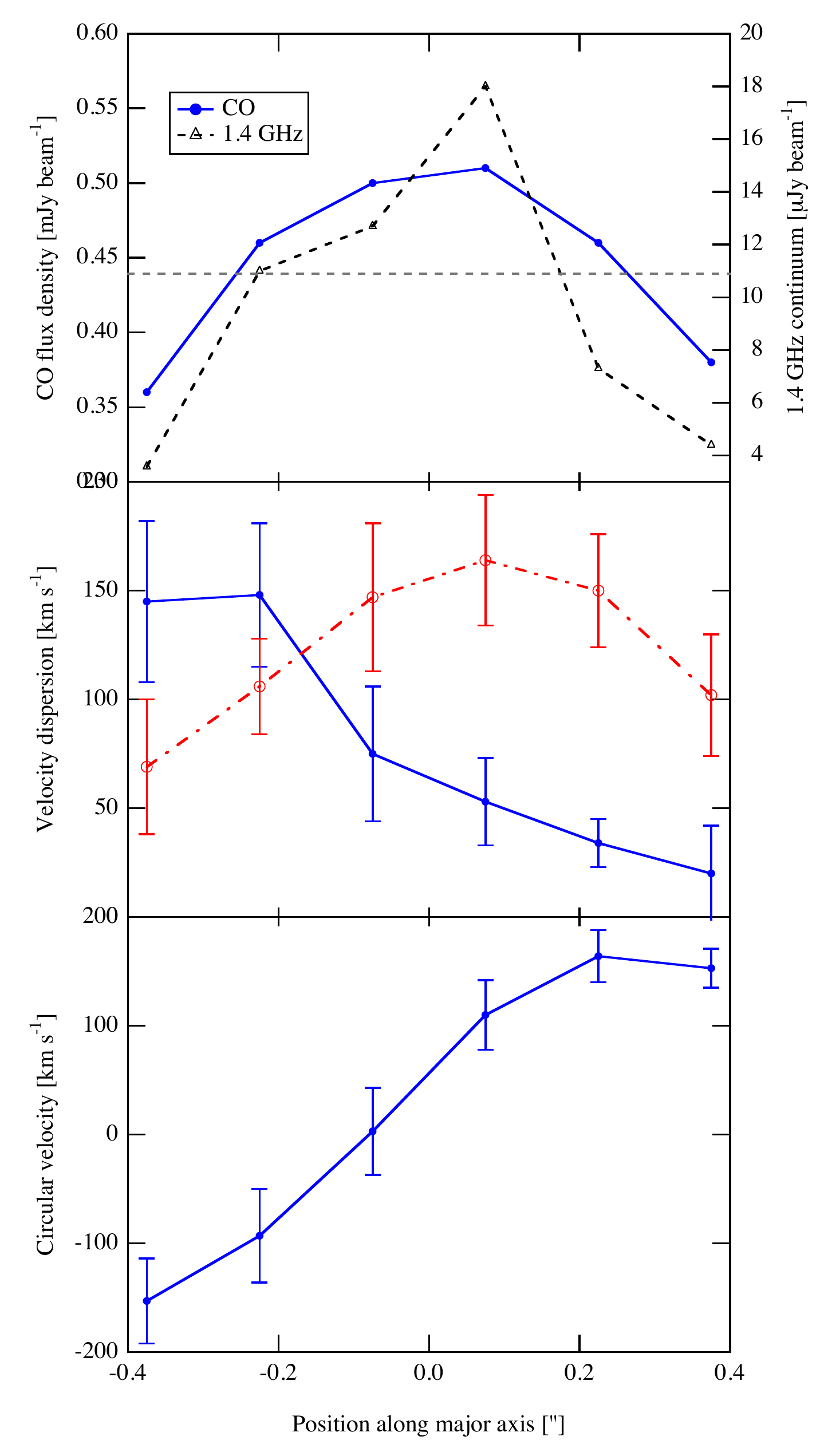}}
\caption{Kinematic analysis for Lockman 38. Spectra were taken at points along the major axis of the galaxy with $\sim$0.15\arcsec\ spacing ($\simeq$ 1/3 - 1/2 of the beam FWHM), and resultant kinematic parameters were derived. \textit{Bottom panel}: circular velocity, obtained using the weighted mean of the two Gaussian peaks fit to the CO emission line. \textit{Middle panel}: velocity dispersion, obtained via the width of the same Gaussian peaks. The solid blue line shows the blue shifted peak, whereas the dashed red line shows the red shifted peak. \textit{Top panel}: CO surface brightness (blue solid line). For reference, the MERLIN-obtained 1.4GHz continuum flux density measured at the same points has been overlaid (black dashed line and right hand axis). The dashed horizontal line is drawn at 11$\mu$Jy/beam ($\sim 3$ times the image rms).}
\label{fig:sfb1_k}
\end{figure}

\subsection{HDF254}
\label{254}

HDF254 is the second brightest (in the radio) of the three sources surveyed. It is also the most extended of the sources, having $>$ 0.5 mJy CO emission extended over nearly 1.5\arcsec\ (= 12.6 kpc at $z=1.999$). It broadly consists of two `knots' of CO emission, an `northern' knot to the north-east, and a `southern' knot to the south west. In addition, there is a slightly weaker CO emission component $\sim$ 0.5\arcsec\ to the west of the southern knot, with a CO flux of 0.67 mJy - this is a strong MERLIN point source (with a peak 1.4 GHz brightness of $\sim$36 $\mu$Jy/beam), which suggests the presence of an AGN. However, the H$\alpha$/NII  observations of \cite{2004ApJ...617...64S}, and PAH features in the NIR spectra taken by \cite{2008ApJ...675.1171P} both show little or no evidence for an AGN. The lower CO knot is detected in the ACS optical image (Fig. \ref{fig:sdb1}, left panel) , and also robustly detected in 1.4 GHz continuum (Fig. \ref{fig:sdb1}, middle panel), implying a high rate of concomitant star formation. By contrast, the northern knot is undetected optically and has a very weak detection in the 1.4 GHz continuum (although radio continuum emission is detected above the $3\sigma$ sensitivity limit of $\sim10 \;\mu$Jy at all points along the source). As the velocity-coded image shows, the two knots are distinct structures in velocity space. Spectra taken across the source show it to be single peaked, with a broad velocity dispersion in the northern knot, and double peaked with lower velocity dispersions in the southern knot. The middle spectrum of Fig. \ref{fig:sdb1} shows a 1-dimensional CO spectrum taken at the position of the MERLIN point source. This shows either continuum emission or a broad peak between -1000 km s$^{-1}$ and -200 km s$^{-1}$ relative to the systemic velocity, at a level of $\sim0.75$ Jy km s$^{-1}$. Integrating the spectrum between these two velocities (where the detection is most robust) gives a signal to noise ratio of $\sim$3. 

The SFR, calculated using integrated low-resolution VLA measurements of the radio continuum, is $1220\pm100$ M$_{\sun}$ yr$^{-1}$. This is possibly somewhat overestimated, however, as the low resolution of the VLA will include the contribution of the possible AGN to the radio flux, artificially boosting the apparent SFR by up to $\sim$1/3 - if the western unresolved MERLIN radio knot is indeed due to an AGN, as expected from the strong hard X-ray emission. \cite{2009MNRAS.399..121C} report a corrected SFR of 1050 M$_{\sun}$ yr$^{-1}$, which we take to be the true SFR hereafter.  

Fig. \ref{fig:sdb1_k} shows a kinematic breakdown for HDF254. All parameters were derived as for HDF132 above. The observed velocity shear of $80\pm28$ km s$^{-1}$ (although again this is a lower limit due to uncertainty in the source inclination) was obtained by fitting a \textit{single peaked} Gaussian profile to the CO emission at points along the major axis of the source. As with HDF132, this method can drastically underestimate the true velocity shear across the system, and in the case of a merging pair may have little physical interpretation (aside from shedding light on the line of sight component of their predominantly transverse velocities). The velocity profile (Fig. \ref{fig:sdb1_k}, bottom panel) resembles a rotation curve, but it may be that the low spatial resolution ($\sim 4$ kpc beam size) of our observations is inadequate to resolve the characteristic signature of a merging pair of bodies. The clumpy (non-disc-like) structure apparent in Fig. \ref{fig:sdb1}, and the fact that the CO morphology changes from being single- to double- peaked as the two separate `knots' are examined, both hint at the source being composed of two distinct bodies which are undergoing the early stages of a major merger. 

The position velocity diagram (Fig. \ref{fig:sdb1}, bottom left) shows the gas motions to be both disordered and complex, which is highly suggestive of a merging system. The two spatially separated bodies are clearly visible, being separated along the position axis and distinct in velocity space. In particular, the southern component clearly consists of two kinematically separate features (manifesting as the double peaked spectrum), whereas the northern component is made up of one single velocity structure.

The MERLIN+VLA radio observations show the star formation (measured using the FWHM of the 1.4 GHz emission map) to be spatially extended over a region $\sim 0.4$ \arcsec\ (= 3.36 kpc) in length. There are five distinct peaks, with a maximum separation of $\sim$1.5\arcsec. The star formation forms an extended structure, and four of the peaks positionally trace the CO emission.  It is, however, much more highly concentrated in the southern, or double peaked knot, as Fig. \ref{fig:sdb1} shows. The top panel of Fig. \ref{fig:sdb1_k} shows this in more detail: while the southern component (right hand peak) is traced very closely by a peak in the radio continuum (with the small 0.1 \arcsec\ offset attributable to astrometrical errors), the northern component (left hand peak) has very little concomitant star formation.

HDF254 has a high velocity dispersion, with the mean value of $\sigma$ being $249\pm18$ km s$^{-1}$. This is far higher than the likely value of $v_{\mathrm{rot}}$, inasmuch as a value of  $v_{\mathrm{rot}}$ can be placed on such a chaotic system. Examining only the southern knot (which due to its double peaked CO spectrum appears to have a rotational component), the value of $\Delta v / 2$ is $149\pm12$ km s$^{-1}$, leading to a ($v_d \sin i/ \sigma)$ of  $0.67\pm0.11$, implying that the southern knot is dispersion dominated, and it likely to be undergoing a major merger. While the northern knot is single peaked, and thus has an undefined rotational velocity, its velocity dispersion is high enough, at $285\pm25$ km s$^{-1}$, to say with some confidence that the northern component too is dispersion-dominated.

Treating the two knots separately it is possible to estimate the mass of the system. The northern knot is a single peaked source, and as such using the peak-to-peak velocity shift as a measure of the Keplerian rotation is not viable. Instead, we adopt a velocity-dispersion approach, as in \cite{2006ApJ...646..107E}, whereby

\begin{equation}
\mathrm{M}_{\mathrm{dyn}} = \frac{C\sigma^2r}{G}
\end{equation}
with the factor $C$ being 3.4 for a rotating disc at an average inclination angle, and 5 for a sphere (see e.g. \citealt{2006ApJ...646..107E}). The mean velocity dispersion of the northern knot region is $264\pm24.8$ km s$^{-1}$. Assuming a $C$ value of 4 (the body is most likely a turbulent ellipsoid, somewhere between the assumptions of a disc and a sphere), the dynamical mass of the northern knot is M$_{\mathrm{dyn}} = 6.9\pm0.4 \times 10^{10}\;\mathrm{M}_{\sun}$.

As explained above, the southern knot has a significant line of sight rotation component, so it is possible to model the component assuming Keplerian motion. The knot has a radius of 0.2\arcsec,which combined with the rotational velocity above gives a dynamical mass for the lower knot of M$_{\mathrm{dyn}}\sin^2 i = 5.6\pm0.3 \times 10^{9}\;\mathrm{M}_{\sun}$, or, using the fiducial inclination of 30\degree\ as above, M$_{\mathrm{dyn}} = 2.2\pm0.2 \times 10^{10}\;\mathrm{M}_{\sun}$. For comparison, it is possible to fit a single Gaussian profile to the spectra and derive the dynamical mass using the velocity dispersion as above. Doing so (the mean velocity dispersion in the lower knot is $201\pm15$ km s$^{-1}$), and adopting a `disc' $C$ value of 3.4, gives a dynamical mass of M$_{\mathrm{dyn}} = 3.1\pm0.2 \times 10^{10}\;\mathrm{M}_{\sun}$. This is comparable (within a factor of two) to the value derived using the peak-to-peak shift: the difference may be due to the poor fitting and artificial broadening of the single Gaussian profile fit to the double peaked spectrum. The total dynamical mass of the entire system, therefore, is M$_{\mathrm{dyn}} = 9.6\pm0.5 \times 10^{10}\;\mathrm{M}_{\sun}$. The dynamical masses of the separate components clearly mark the system as being a major merger, with a mass ratio of (2.2 - 3.1):1. 

To calculate surface densities, we integrate values across the surface of the entire elongated body of the source. As for HDF132, we take the total gas mass from a low-resolution observation of the same source, to avoid resolving out flux. \cite{2009arXiv0910.5756C} report the total gas mass of this source to be $8.3 \times 10^9$ M$_{\sun}$, suggesting that our high-resolution observations resolve out a factor of 2 in CO flux. Based on the CO morphology, we assign source semi-major and semi-minor axes (measured using the half light radius as explained above) of 0.52\arcsec $\times$ 0.16\arcsec, which leads to a gas surface density value of $\Sigma_{\mathrm{H2}}$ = 513 M$_{\sun}$ pc$^{-2}$. Adopting an identical method for the SF surface area, we fit the radio emission area with semi-major and -minor axes of 0.41\arcsec\ $\times$ 0.14\arcsec. This gives a SFR surface density of  $\Sigma_{\mathrm{SFR}}$ = 58 M$_{\sun}$ yr $^{-1}$  kpc$^{-2}$.


\subsection{Lockman 38}
\label{38}
Lockman 38 is the weakest radio source in our sample, suggesting that it represents an example of the extreme radio-faint end of the SMG population, having a SFR typical of the faintest 20\% of SMGs as presented by \citealt{2007MNRAS.380..199I}. This is similar to the brightest  BX/BzK galaxies at $z\sim1.5$.

The CO emission is moderately extended in the north-south direction, with the strongest emission coming from an elliptical region 1.1\arcsec\ in diameter, 9.3 kpc at $z=1.523$. It is comprised of two main velocity components, a small blue-shifted region to the north and a larger (and brighter) red-shifted region to the south. As with HDF132, there is a fainter (by a factor of $\sim$2) `tail' emission component, with a velocity intermediate between the two `main' regions. 1D spectra taken at positions along the source show it to be double peaked at all points. As with HDF132, the position of each respective peak is close to constant, with the position of the red-shifted peak varying by $\pm 33$ km s$^{-1}$ along the source, and the position of the `blue' peak varying by $\pm 38$ km s$^{-1}$. Unlike HDF132, however, the peaks are highly asymmetric, with the red-shifted peak near $v\sim0$ km s$^{-1}$ being around a factor of 5 brighter than the blue-shifted peak at $-400$ km s$^{-1}$.

The position-velocity diagram shows the gas motion to be somewhat disturbed. While there is no evidence for spatially separated bodies, there are two distinct kinematic components, which are separated by $v\sim-400$ km s$^{-1}$ in velocity space. These peaks are asymmetric (as discussed above), and do not display the signature of a disc-component. The position-velocity diagram suggests a late-stage merger interpretation of Lockman 38 might be most appropriate, where the components have coalesced to form a single body, but are still kinematically distinct.

The MERLIN+VLA radio continuum map shows the 1.4 GHz emission to be strongly peaked within the brightest region of the galaxy, with little or no detected emission in either the blue-shifted region to the north or the fainter tail. A spatial breakdown of this (Fig. \ref{fig:sfb1_k}, top panel) shows the radio emission falling below the sensitivity threshold of 10$\mu$Jy for all but the inner $\sim 0.3$\arcsec\ of the source.

A kinematic investigation of the source (Fig. \ref{fig:sfb1_k}) shows it to have the largest velocity gradient of the three sources (bottom panel), having a maximum $v_{\mathrm{rot}}$, obtained as above using a weighted mean of both peaks, of $164\pm24$ km s$^{-1}$. This velocity varies smoothly across the source. Using the peak-to-peak shift ($\Delta v / 2$), a better measure of the true maximal rotation velocity, $v_{\mathrm{rot}} = 229\pm28$ km s$^{-1}$.  The velocity dispersion for the two separate peaks is shown in Fig. \ref{fig:sfb1_k}. The value of $\sigma$ for both peaks varies along the source, though in opposite directions. The mean value for the dispersion is $79\pm22$  km s$^{-1}$ and $123\pm24$ km s$^{-1}$ for the blue-shifted and red-shifted peaks respectively. The value of $v_d (\sin i)/ \sigma$ for this source (taking a mean of the two peak dispersions) is $2.3\pm0.14$, placing the source very much in the rotation-dominated regime. However, due to the disordered and non-rotational gas motions revealed in the position-velocity diagram, we class this source as a late-stage merger, as it is not dispersion-dominated, and apparently non-clumpy on the scales revealed by our beam, yet does not seem to behave kinematically as a true disc. 

As described in \S \ref{gannon} above, Lockman 38 has never been observed in the compact configuration required to measure a truly accurate gas mass - we can, however, estimate the true value by assuming that with our high-resolution observations we resolve out a similar fraction to the other sources in our programme. We therefore assume that we lose a factor of 1.9 in flux, and adjust our derived gas surface density accordingly. 

\cite{2009MNRAS.397..281I} VLA imaging of Lockman 38 reveals a total flux density of $32.3\pm7.2$ $\mu$Jy, if the source size is assumed to be identical to that of the beam. However, the source is faintly resolved (vaguely consistent with a bipolar morphology) even by the 3.9" beam. If the source size is deconstrained, the total flux density is $56.2\pm22.4$ $\mu$Jy, which is taken to be the true 1.4 GHz flux hereafter. This large radio extent (resolved by the large beam) suggests that in addition to the compact core there is a low-level extended starburst being fuelled by the extended CO reservoir. The 610 MHz flux density, obtained by \cite{2009MNRAS.397..281I} with the GMRT is $59.8\pm17.9$ $\mu$Jy, which supports the model of this system as a starburst (as opposed to a steep-spectrum AGN).

Lockman 38 has an unusually low radio-derived star formation rate for an SMG, at $250\pm80$ M$_{\sun}$ yr$^{-1}$. This is despite having a sub-mm flux of $6.5\pm1.9$ mJy, implying a SFR of $\sim$600 M$_{\sun}$ yr$^{-1}$. This low radio/sub-mm ratio is unusual given its redshift, and hints at an unusually cold dust temperature for this source. Assuming the SFR(radio) to be correct, we again calculate gas and SF surface densities for the source. Due to the highly asymmetric nature of the peaks, we do not take a `mean' value of L$^{\prime}$(CO), but instead use the more significantly detected stronger peak. We fit the significant (R$_{1/2}$) CO emission with an ellipse, with semi-major and semi-minor axes of 0.31\arcsec\ and 0.56\arcsec respectively. This gives a gas surface density (including the estimated factor of 1.9 to account for flux resolved out) of $\Sigma_{\mathrm{H2}}$ = 300 M$_{\sun}$ pc$^{-2}$. The radio continuum emission is more highly concentrated than the CO emission, and is well fit by a 0.29\arcsec\ $\times$ 0.25\arcsec\ ellipse; using the SFR given in Table \ref{tab2}, we arrive at a SFR surface density of  $\Sigma_{\mathrm{SFR}}$ = 14 M$_{\sun}$ yr $^{-1}$  kpc$^{-2}$.

To estimate the dynamical mass of Lockman 38, we assume rotational motion, and use the peak-to-peak velocity shift as a measure of its maximal rotation velocity; we take the radius to be 0.56\arcsec (= 4.75 kpc) and the rotation velocity as $229 \pm 28$ km s$^{-1}$. Assuming Keplerian motion, we calculate the dynamical mass to be M$_{\mathrm{dyn}}$ $\sin^2 i =1.6 \pm 0.9\times 10^{10}$ M$_{\sun}$. Taking the fiducial inclination of 30\degree, the dynamical mass is calculated to be M$_{\mathrm{dyn}}$ = $6.2 \pm 1.1\times 10^{10}$ M$_{\sun}$.

\subsection{HDF255}
\label{255}

HDF255 falls within the field of view for the HDF254 observation, being just $\sim$7\arcsec\ to the south east. It is, however, undetected by our long-baseline `A' configuration observations, suffering from both resolution-based flux loss in the extended source, and primary beam attenuation effects (being off phase centre). Figure \ref{fig:255} shows $^{12}$CO(J=4$\rightarrow$3) observations taken with the low-resolution, compact `D' configuration of the PdBI (obtained for our own pilot observations for this programme), along with stacked $b, v, i$ \& $z$ band optical imaging from the ACS.

HDF255 has been observed in the radio continuum with both MERLIN and the VLA, so calculating the star formation rate surface density is possible. The FWHM of the MERLIN+VLA image has an elliptical size of 0.56\arcsec\ $\times$ 0.32\arcsec, and the 1.4 GHz flux density  observed from the VLA is $53.9\pm8.1\mu$Jy. K-correcting and applying the radio-far IR correlation (as per Eq. 1), this gives a SFR of $621 \pm 93 \,\mathrm{M}_{\sun}\, \mathrm{yr}^{-1}$, with a resultant SFR surface density of $\Sigma_{\mathrm{SFR}} = 16 \; \mathrm{M}_{\sun} \,\mathrm{yr}^{-1} \,\mathrm{kpc}^{-2}$.

The galaxy is well detected in CO ($\sim 4.5 \sigma$) by the compact configuration, but is poorly detected by the long baseline observations, suggesting that it is a low surface density source with a large extent which has been driven below the sensitivity threshold by the small beam (0.4\arcsec) of the `A' configuration.  The compact-configuration flux for HDF255 is S$_{\mathrm{CO4-3}} = 0.53 \pm 0.11$ Jy km/s, which leads to a gas mass of $3.5\pm0.7 \times 10^9 \,\mathrm{M}_{\sun}$ (after correction to an L\arcmin$_{\mathrm{CO(1-0)}}$ equivalent as above). This relatively low gas mass, coupled with a source size larger than the `A'-configuration beam at 154 GHz, leads to the source being resolved out (we would only detect this were all the flux to be concentrated within a beam size). We assume an approximate (lower limit) CO source radius of 0.6\arcsec, based on the high-resolution radio source size and the extent of the beam. This is a rough approximation at best, however, and should be treated with caution. This leads to a value of $\Sigma_{\mathrm{gas}} \leq 125\; \mathrm{M}_{\sun} \,\mathrm{pc}^{-2}$.

\section{Discussion}
 
 \begin{table*}
\centering
\caption{Properties of all SMGs, both observed in this programme and archival. All parameters for archival sources have been re-calibrated with the same conversion factors as given in this work. Half light radii for our sources are approximations to our (non-circular) sources for comparison only - see Table 2 for accurate source dimensions, which were used to derive all parameter surface densities.}
\begin{tabular}{ccccccc}
\hline
\hline
Source  &z& R$_{1/2}$ (CO) & R$_{1/2}$ (1.4 GHz) & log $\Sigma_{\mathrm{gas}}$& log $\Sigma_{\mathrm{SFR}}$ & MERLIN data reference\\
& &[kpc]&[kpc] &[M$_{\sun}$ pc$^{-2}$] &[M$_{\sun}$ yr$^{-1}$ kpc$^{-2}$]\\
\hline
\hline
\textbf{Our sources:}\\
HDF132 & 1.99 &2.7&1.8&2.73&2.16 & --\\
HDF254 & 1.99 & 2.9 & 2.3 & 2.71 & 1.76 & --\\
Lockman 38 & 1.52 & 3.7 & 2.3 & 2.48 & 0.76 & -- \\ \\

\textbf{Archival sources:}\\
SMMJ105141+571952 & 1.21 & 3.0 & 2.7 & 2.59 & 1.45 & Unpublished\\
SMMJ123634+621241 & 1.22 & 4.1 & 4.7 & 2.51 & 0.83& Unpublished\\
SMMJ123549+6215 & 2.20 & 0.9 & 1.0 & 3.87 & 2.45& Unpublished\\
SMMJ123707+6214w & 2.49 & 2.8 & 3.5 & 2.59 & 1.12 & \cite{2004ApJ...611..732C}\\
HDF255 & 1.99 & $\geq 5$ & 3.2 & 2.13 & 1.19 &\cite{2004ApJ...611..732C}\\
\hline
\end{tabular}
\label{tab3}
\end{table*}

\subsection{Mergers vs. discs}

It is challenging to distinguish between disc-type structures and late stage merger remnants at high redshift (e.g. \citealt{2008ApJ...680..232D}; \citealt{2008AJ....136.1110M}). Indeed, SMGs are generally classed as either undergoing a major merger, or consisting of a compact rotating merger remnant (\citealt{2006MNRAS.368.1631S}; \citealt{2008ApJ...680..246T}; \citealt{2008ApJ...682..231S}). In our sample, certainly HDF254 is a system undergoing an early stage major merger: as explained above in \S\ref{254}, the galaxy appears clumpy on sub-arcsecond scales, and has an extremely high velocity dispersion (see Fig. \ref{fig:sdb1_k}), indicative of the chaotic and disordered gas motions created by such a merger. 

HDF132, however, does not appear to be undergoing an early stage merging event; its symmetrical velocity peaks, low velocity dispersion (relative to the rotational velocity), and ordered gas motions lend some credence to the non-merging picture. The NICMOS IR morphology of HDF132 reveals an elliptical structure, with the major axis coincident (within the astrometrical errors) with the major axis of the CO emission. If we are to class HDF132 (and potentially Lockman 38) as a system not undergoing a major merger, or consisting of a recent merger remnant, we must be able to place these systems into a valid, cosmologically-motivated paradigm of galaxy formation which allows for extreme star formation rates and extended disc-like structures to emerge from hot/cold flow accretion type models.

Recent simulations of giant galaxies at high redshifts of $z\sim2$ (\citealt{2009arXiv0909.4078D}; \citealt{2009Natur.457..451D}) have suggested that extended CO discs can form entirely secularly, without a major-merging event. In this `smooth accretion' model, the systems formed are `harassed' massive galaxies, which are being rapidly formed by filamentary accretion of material from the IGM. They can be pictured as undergoing a series of minor mergers (with dark matter halo ratios of $>$1:10), which are observationally indistinguishable from a smooth-accretion driven secular evolution. 

\cite{2009arXiv0905.2184N} carried out simulations of the star forming molecular gas within SMGs, using the \textit{a priori} assumption that SMGs are merger-driven systems. The velocity field (Fig. 6 of their paper) shows, in general, a chaotic and disordered structure at all times apart from a short period ($\sim 0.03$ Gyr) in the late inspiral phase of the merger; a system at this stage is yet to undergo the `final coalescence' into a compact merger remnant. It is this final coalescence that boosts the star formation rate and 850$\mu$m flux, which causes the system to be classed as an SMG, or more generally a ULIRG. The reasonably bright 850$\mu$m flux ($>7$ mJy) and high star formation rate ($>10^3$ M$_{\sun}$ yr$^{-1}$) argue against HDF132 being an inspiralling merger, suggesting instead that the final coalescence has already occurred. This conflict lends weight - at least from a simulation-led perspective - to the argument that the behaviour of HDF132 cannot be easily explained by a merger. It must be noted, however, that while these simulations do manage to reproduce the high star formation rates seen in SMGs ($>10^3 \;\mathrm{M}_{\sun} \;\mathrm{yr}^{-1}$) from a hydrodynamical perspective, the explicit model employed is unable to account for the features of the entire SMG population, and as such must be treated with care when dealing with individual cases. 

\subsection{Size of CO emission regions}

Recent work investigating the sizes of massive high redshift galaxies has shown that some high-$z$ galaxies do, in fact, possess extended morphologies. For example, \cite{2009arXiv0909.3088M} investigated a sample of massive early type galaxies, finding that their optical radii were comparable to the radii of similar mass galaxies in the $z=0$ Universe, suggesting that the low S/N of massive high-$z$ galaxies causes their half-light radii to be systematically underestimated. 

HDF254, the only unequivocally merging system in our study, is comprised of two or three smaller bodies, each $\sim$ 2 kpc in diameter, comparable in extent to the mean size of SMGs identified by (for example) \cite{2008ApJ...680..246T}. This is the first size measurement for a sub-mm faint ULIRG. However,  the total system size is $>12$kpc, and unlike other comparable mergers from \cite{2006ApJ...640..228T} and \cite{2008ApJ...680..246T} such as HDF242, and 0931, there is a `bridge' of CO emission connecting the interacting clumps. Both HDF132 and Lockman 38, however, have CO spatial extents larger than typically seen in high resolution studies of SMGs. HDF132 is the most compact of our sources, but with a diameter of $\sim 0.7$\arcsec\ ($= 5.9$ kpc) is more extended than all but one of the 11 sources presented by \cite{2006ApJ...640..228T}. Lockman 38 has a diameter of 1.1\arcsec\, (9.5 kpc at $z=1.523$). The prevailing paradigm of SMGs being major mergers leads naturally to the conclusion that their molecular gas will be concentrated in a compact nuclear region: tidal torquing induced by the merger drives cold molecular gas into the centre of the galaxy, where it is heated, becomes gravitationally unstable, and triggers a starbursting event.  This suggests that while systems undergoing major mergers may comprise a part of the SMG population, it is perhaps not exclusively so.


\subsection{The L\arcmin$_{\mathrm{CO}}$-L$_{\mathrm{FIR}}$ Relation and Star Formation Efficiency} 

A common way to conceptualise the relationship between the gas content and star formation rate of a galaxy is in terms of a star formation efficiency, defined as the SFR per unit of star forming gas - i.e. SFR/M(H$_2$). Measuring this for high-$z$ galaxies requires an assumption of a CO-H$_2$ conversion factor; while this is a necessary evil for calculating gas masses and gas surface densities, it is possible to frame the star formation efficiency in terms of observables (assuming the FIR-radio correlation), and use the continuum-to-line luminosity ratio L$_{\mathrm{FIR}}$/L\arcmin$_{\mathrm{CO}}$.

\begin{figure}
\centerline{\includegraphics[scale=0.53]{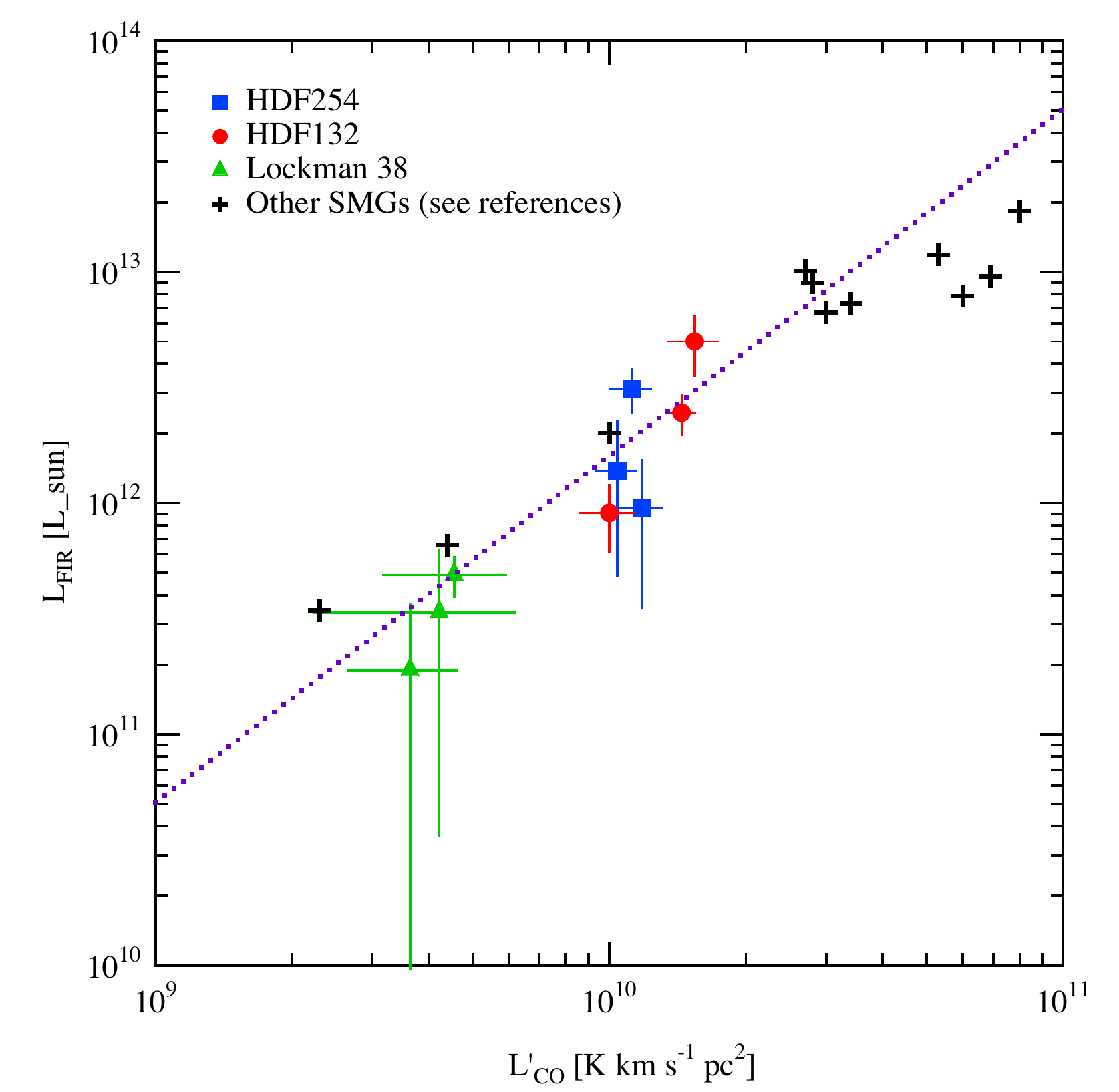}}
\caption{CO luminosity plotted against FIR luminosity for `sub-clumps' of our three sources. For reference, results for other SMGs in the literature have been plotted (\citealt{2003ApJ...584..633G}; \citealt{2003ApJ...597L.113N}; \citealt{2003ApJ...582...37D}; \citealt{2004ApJ...614L...5S}; \citealt{2005MNRAS.359.1165G}; \citealt{2009ApJ...695L.176D}; \citealt{2009A&A...496...45K}). The dashed line represents the fit to SMGs and ULIRGs from Greve et al. (2005) of (log L\arcmin$_{\mathrm{CO}} = 0.62 (\log$ L$_{\mathrm{FIR}}) + 2.33$). All our sources scatter around the canonical relation by up to a factor of 5.}
\label{fig:sfe}
\end{figure}

We have examined this continuum-to-line luminosity ratio for three `sub-regions' of each of our sources, to better show the star forming behaviour within each. Fig. \ref{fig:sfe} shows the result for each of the galaxies. For comparison, other global values of L$_{\mathrm{FIR}}$/L\arcmin$_{\mathrm{CO}}$ for SMGs in the literature have been plotted - see the caption for references. All the sub-clumps scatter (by a factor of $\sim5$) around the canonical L$_{\mathrm{FIR}}$/L\arcmin$_{\mathrm{CO}}$ relation as defined by \cite{2005MNRAS.359.1165G}. This relation has a slope less than unity, indicating that as L$_{\mathrm{FIR}}$ increases, as does the L$_{\mathrm{FIR}}$/L\arcmin$_{\mathrm{CO}}$ ratio: star formation is more efficient in ultraluminous systems like SMGs than in LIRGS and normal star forming galaxies. However, \cite{2005ApJ...630..250K} have suggested that it is not a star formation efficiency issue, but rather results from a disconnect between the gas being traced and the gas actively involved in star formation (see \citealt{2007ApJ...660L..93G}, for example). In this picture, CO emission with J$_{\mathrm{upper}} \leq 4$ traces the ambient gas density (which is non-linear with SFR), whereas emission from molecules that trace the densest star forming cores (such as HCN) is better suited for measuring the density of the gas actually involved in star formation. Indeed, in the local Universe L(HCN) is seen to increase linearly with L(FIR). The sub-clumps of our three SMGs follow this (non-linear) trend, and have L$_{\mathrm{FIR}}$/L\arcmin$_{\mathrm{CO}}$ ratios consistent with other SMGs. As can be seen from Fig.  \ref{fig:sfe}, Lockman 38 is the least-efficient SMG studied to date, further suggesting that it is a highly unusual system.

\subsection{Gas fractions and the uncertain CO conversion factor}

\begin{figure}
\centerline{\includegraphics[scale=0.5]{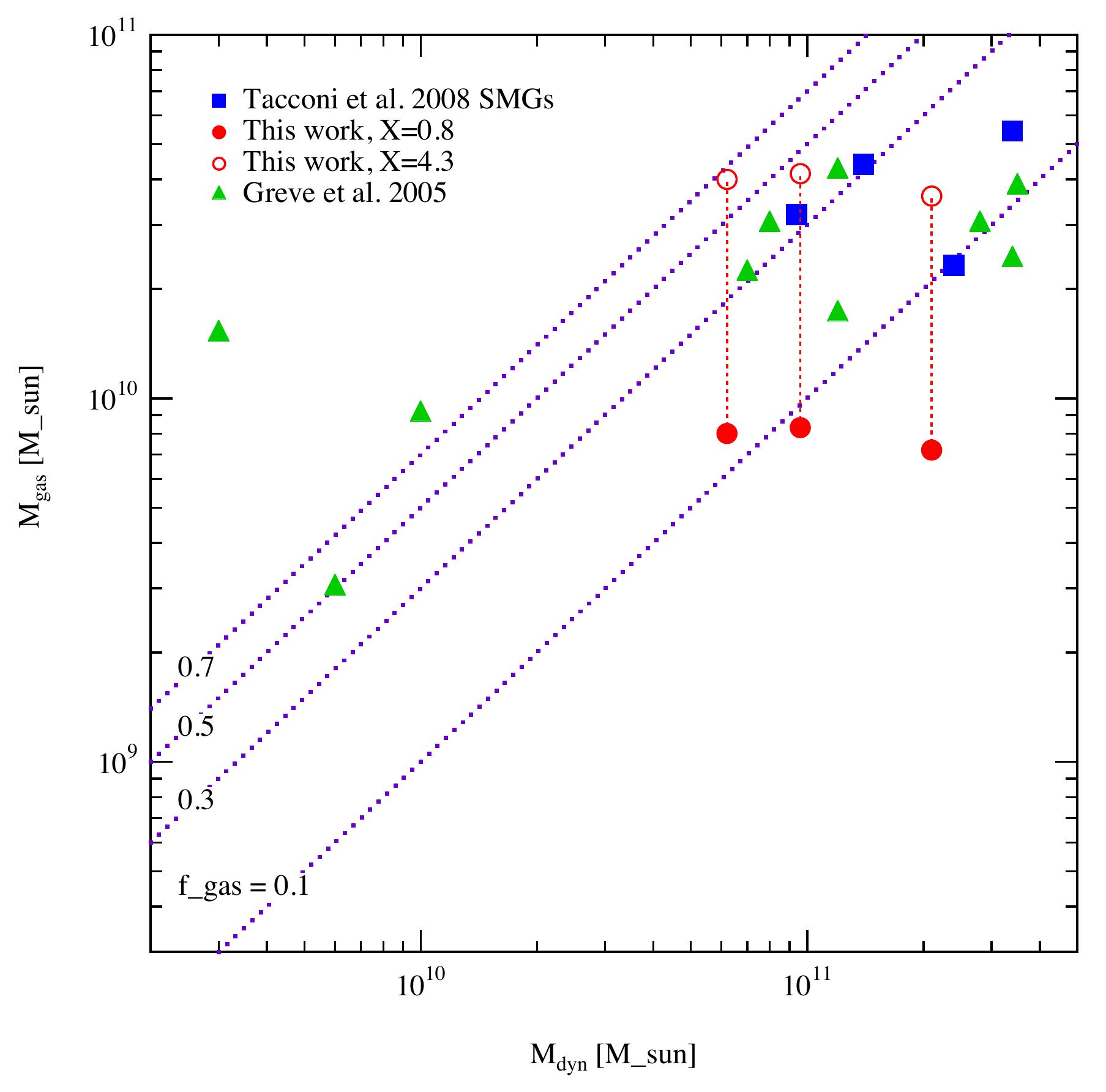}}
\caption{Dynamical mass plotted against molecular gas mass for all sources presented in this work (filled red circles). For comparison, high-$z$ star forming galaxies and SMGs from Tacconi et al. (2008) and Greve et al. (2005) are also shown. Open red circles represent the position of the three sources presented in this work, if a Milky Way CO conversion factor is used to calculate the gas mass. Dashed diagonal lines, from bottom to top, represent constant gas fractions (i.e. M$_{\mathrm{gas}}$/M$_{\mathrm{dynamical}}$) of 0.1, 0.3, 0.5 and 0.7 respectively.}
\label{fig:fgas}
\end{figure}

The question of the appropriate CO/H$_2$ conversion factor to use in extreme high-$z$ systems is an ongoing one. The generally accepted values for local ULIRGs are within the range $0.8 - 1.6\; \mathrm{M}_{\sun}\; \mathrm{K  \;  km\; s}^{-1} \mathrm{pc}^{2}$, substantially lower than the calibration used for `normal' systems - this value (normally referred to as the Milky Way, or galactic, conversion factor) is generally taken to be $4 - 5\; \mathrm{M}_{\sun}\; \mathrm{K  \;  km\; s}^{-1} \mathrm{pc}^{2}$, and is derived both theoretically and empirically (see Appendix A of \citealt{2008ApJ...680..246T} and references therein for discussion). Obtaining the correct value for any given system is difficult, however, as the true value is a complex function of many physical parameters, including the gas metallically \citep{2005A&A...438..855I}, and the column density and temperature of the intersteller gas (\citealt{1996ApJ...462..215S}; \citealt{2001A&A...365..571W}; \citealt{2007MNRAS.379..674W}). 

The theoretical justification for using a significantly lower conversion factor for ULIRGs than normal star-forming galaxies hinges on the nature of merger-induced star formation. Whereas `normal' star formation occurs in gravitationally bound molecular clouds (which exist as well defined separate structures in the ISM), the torque induced by a merger event creates a more centralised, homogeneous molecular ISM, in which pressure plays a significant role in triggering a star forming event. This difference (gravitationally-confined vs. pressure confined star forming gas) theoretically justifies the use of a different conversion factor. Empirically, it is found that using a standard galactic conversion factor for ULIRGs often results in derived gas masses greater than their dynamical masses, necessitating the adoption of a lower value. 

The paradigm of SMGs being `scaled up' versions of local ULIRGs has led to these lower values of $X_{\mathrm{CO}}$ being adopted by most authors (i.e. $X=0.8$ by \citealt{2005MNRAS.359.1165G}; $X=1$ by \citealt{2008ApJ...680..246T}), which has led to a predominance of low assumed gas fractions for SMGs, which typically have values of f$_{\mathrm{gas}}$ (= M$_{\mathrm{gas}}$/M$_{\mathrm{dynamical}}$) between 0.1 and 0.3. Fig. \ref{fig:fgas} shows the relationship between dynamical mass and gas mass for the sources in this work, along with a selection of SMGs from the literature - diagonal lines of constant gas fraction have been drawn for clarity. A majority of the galaxies shown (9/17) have low gas fractions of $<20\%$, a value which would be modest for local star forming discs (e.g. \citealt{2004AJ....127.2031K}; \citealt{2009MNRAS.400..154B}) and seems utterly incongruous with the exceptional star formation rates achieved by SMGs.

For the 3 ULIRGs considered in this paper, we have measured relatively extended CO structures of low gas surface density, comparable to local starburst galaxies, and distinctly different from the dense, nucleated gas of local ULIRGs. It is therefore reasonable to consider that the use of the CO conversion factor appropriate for local ULIRGs ($X \sim 1$) might not be justified in the types of systems uncovered in this work. We also show in Fig. \ref{fig:fgas} (as open red circles) the effect of using a Milky Way conversion factor ($X = 4.3\; \mathrm{M}_{\sun}\; \mathrm{K  \;  km\; s}^{-1} \mathrm{pc}^{2}$) for our three galaxies. 

A full theoretical and empirical treatment of the CO/H$_2$ conversion factor lies beyond the scope of this work. We do, however, include in the following discussion of the star formation law the possibility that the assumed value of $X = 0.8\; \mathrm{M}_{\sun}\; \mathrm{K  \;  km\; s}^{-1} \mathrm{pc}^{2}$ may be significantly underestimating the true gas mass.


\subsection{The Star Formation Law}

There have been many recent studies on star formation at high redshifts, putting the extreme systems seen at $z\sim2$ (where the cosmic star formation rate density peaks) into the context of a `universal star formation law'. The most common prescription, at both low and high redshifts, is the power-law scaling between the surface density of gas ($\Sigma_{\mathrm{gas}}$) and star formation ($\Sigma_{\mathrm{SFR}}$) known as the Kennicutt-Schmidt (KS) law  (\citealt{1959ApJ...129..243S}; \citealt{1998ApJ...498..541K}). 

A full and correct calibration of the high-$z$ KS law is crucial in order to facilitate studies of star forming galaxies at cosmological redshifts. With current instrumentation it is only possible to measure CO masses for the most luminous systems: this has the result that all galaxies currently observed in CO at $z > 2$ are ULIRGs, which are clearly unrepresentative of the high-$z$ population as a whole. In addition, obtaining accurate gas masses via CO measurements is highly time consuming, requiring on-source integration times of several hours; as such, obtaining well constrained gas masses for a statistically significant sample of high-$z$ star forming galaxies would be prohibitively expensive in terms of instrument time. One way around this is to infer a gas mass based on the (more easily measurable) H$\alpha$ emission size and star formation rate (e.g. \citealt{2006ApJ...647..128E}). This requires the assumption of a universal star formation law, which has only been truly calibrated at low redshifts, and may be inaccurate for both the `normal' (L$^*$ galaxies) and extreme systems (SMGs, SFRGs etc.) lying at cosmological redshifts. Clearly, for this kind of large scale work to be undertaken there is a need for high redshift galaxies to be placed into the context of a KS-type star formation law.  

\begin{figure}
\centerline{\includegraphics[scale=0.53]{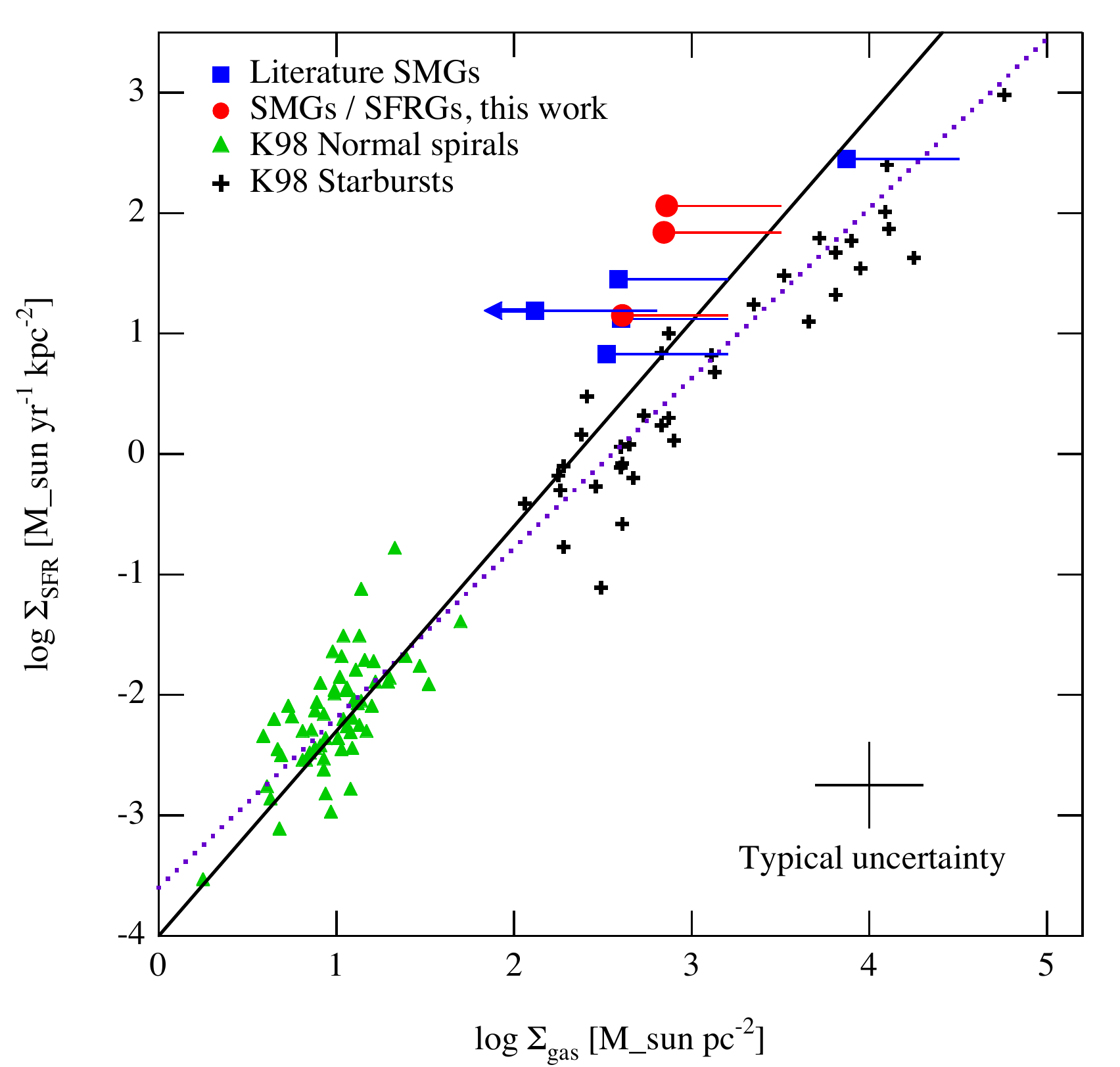}}
\caption{Log gas surface density plotted against log star formation surface density (the `Kennicutt-Schmidt relation'; see \citealt{1998ApJ...498..541K}). Black crosses and green triangles represent the normal spirals and circumnuclear starbursts (respectively) published by Kennicutt (1998b). Blue squares denote previously derived values of $\Sigma_{\mathrm{gas}}$ and  $\Sigma_{\mathrm{SFR}}$ for SMGs in the literature, and red circles denote the same parameters for the three new galaxies presented in this work. Previously published SMG results have been adjusted to a common CO/H$_2$ conversion factor of $0.8\; \mathrm{M}_{\sun}\; \mathrm{K  \;  km\; s}^{-1} \mathrm{pc}^{2}$, and all SMG molecular gas masses have had a helium correction factor of 1.36 applied. Horizontal `error bars' on the SMGs/SFRGs extending to the right show the range of gas surface densities achievable by varying the CO/H$_2$ conversion factor, up to a Milky Way value of 4.3. The dashed line is the original ($N=1.4$) slope given by Kennicutt (1998b), and the solid fit is the improved fit ($N=1.71$) applicable to high-$z$ galaxies as presented by Bouch\'e et al. (2007).}
\label{fig:ks}
\end{figure}

\cite{2006ApJ...647..128E} established a plausible KS-law relation for BX/BM galaxies by inverting the KS law for gas surface densities.  \cite{2006ApJ...640..228T} explicitly measured the gas surface densities in SMGs, and inferred the basic KS-law properties of SMGs, while \cite{2007ApJ...671..303B}  reviewed the literature on this topic, and placed all these galaxies into a self-consistent framework, re-iterating the findings of both these authors. \cite{2007ApJ...671..303B} report that, in general, extreme star forming galaxies at high redshifts lie above the $z\sim0$ KS law originally proposed by \cite{1998ApJ...498..541K}, a fact which they attribute to an environmental enhancement of the star formation rate most likely due to collisions or pressure. When including the high-$z$ galaxies, they modify the canonical KS slope of $N\sim1.4$, finding that $N=1.71\pm0.05$ provides a better fit. Even taking this steeper relation into account, the SMGs in their study still lie above the relation, indicating the very high efficiency of their star formation. \cite{2007ApJ...671..303B} were, however, forced to make assumptions, including the fact that the sizes of the CO region and the star forming region were identical, using the half-light radius (R$_{1/2}$) of the CO emission for both.

Our high-resolution, matched beam study of SMGs allows us to improve over previous studies by taking independent observations of both the sizes of the star forming and the CO emission regions, and by utilising accurate and independent measures of star formation rates and gas masses. As discussed in \S3 above (and see Table \ref{tab2}), we find that in all instances the star forming region is centrally peaked, and is extended over a smaller region than the CO morphology. Taking the `size' of the respective emission regions to be the half light radius, we calculate the surface density of both star formation and molecular gas (which we correct for interstellar helium by multiplying our molecular gas masses by 1.36). 

Fig. \ref{fig:ks} shows our three sources in the context of the KS law. The smaller size of the 1.4 GHz emission region (respective to the CO emission region) leads to the three sources presented in this work lying significantly above the $z\sim0$ KS relation, and above the high-$z$ ULIRGS previously examined. This indicates an extremely high efficiency of star formation respective to the gas density. Even Lockman 38 - which may be forming stars an order of magnitude slower than a typical SMG (as per the discussion above) - has such an extended CO morphology and compact radio emission that its gas surface density is lower than would be expected given its SFR, causing it to lie above the high-$z$ KS relation. Both HDF132 and HDF254 lie significantly ($\sim 0.8$ dex) off the KS slope: it is likely to be a combination of the extended CO morphology and compact radio emission which conspire to move the sources above the relation. As discussed above however, this finding is dependent on the assumption of a merger-appropriate CO/H$_2$ conversion factor for our sources, which may be significantly underestimating the gas content for our galaxies, particularly the disc-like HDF132 which has a gas fraction of 5\% assuming a ULIRG-appropriate conversion factor. If the smaller conversion factor is rejected and instead a Milky Way factor is used, then our galaxies fall very close to the high-$z$ KS relation of \cite{2007ApJ...671..303B}.  

Using our independent measurements of the MERLIN-observed star forming regions and the CO morphology, we have also re-calculated $\Sigma_{\mathrm{gas}}$ and $\Sigma_{\mathrm{SFR}}$ for SMGs from the literature: these properties are presented in Table \ref{tab3}. For consistency, all gas surface densities have been adjusted to a common CO/H$_2$ conversion factor of $0.8\; \mathrm{M}_{\sun}\; \mathrm{K  \;  km\; s}^{-1} \mathrm{pc}^{2}$ (though an error bar again delineates the possible range of values, up to the standard Milky Way value) , and all SMG molecular gas masses have had a helium correction factor of 1.36 applied. These archival SMGs are also shown in Fig. \ref{fig:ks}. There appears to be an offset between the SMGs presented in this work (both newly observed and archival) and the KS relation for high-$z$ star forming galaxies claimed by \cite{2007ApJ...671..303B}, which suggests that obtaining an estimate of a galaxy's gas mass by inverting the canonical KS law can give unreliable results. Furthermore, as explained above the offset between our sources and other suggested KS law slopes is, perhaps, partly attributable to a combination of uncertainties in the CO/H$_2$ conversion factor and star formation rates for this type of system, but could also provide evidence that stars are formed more efficiently in these extreme systems. If this is the case, then it may be that a `universal star formation law', which provides a consistent power-law relationship between gas content and star formation across all systems and environments, is not achievable, or at least does not apply in the most extreme cases. 

\section{Conclusions}

We have presented high resolution (sub-arcsecond) detections of molecular gas in three $z\sim2$ star forming galaxies, two SMGs and one SFRG. We used extracted spectra to constrain CO fluxes, and extrapolated total molecular gas masses. We analysed the kinematic behaviour of the sources revealed in the CO imaging on 0.15\arcsec\ scales, equal to $\sim 1/3 - 1/2$ of the beam FWHM. Our main conclusions are as follows:

\begin{itemize}
\item{One of our galaxies is unequivocally an early stage major merger (HDF254) with high velocity dispersion and a minimal rotational contribution. Lockman 38 appears to be a late stage merger, with a low velocity dispersion and a smooth morphology, but with significantly disturbed gas kinematics revealed by the position-velocity diagram. The remaining detected source (HDF132) appears - to the limits of our data - consistent with being a disc-like rotating body, with double peaked profiles indicative of rotation and a homogeneous, non-clumpy morphology. These features may be explainable with a late stage merger model, or may be evidence of a star forming disc.}
\item{There is a marked discrepancy between the sizes of the CO and 1.4 GHz radio emission regions in our high-resolution data, with the radio emission region (which traces star formation) being more compact than the CO region in all cases. This has important implications for our understanding of star formation in high-$z$ starburst galaxies, particularly in relation to the role of gas reservoirs, and whether starburst galaxies tend to be fuelled by cold flow accretion, or minor mergers (see \citealt{2006MNRAS.368....2D}).}
\item{This size difference leads to an exacerbation of the offset of SMGs from the Kennicutt-Schmidt star formation law, which has already been shown to be significant (see \citealt{2007ApJ...671..303B} for a compilation of recent findings in this area).}
\item{The adoption of a low, ULIRG-like CO/H$_2$ conversion factor for SMGs leads to very low gas fractions of $<30\%$, which is far lower than would be expected given their extreme SFR. The adoption of a Milky Way conversion factor (which is somewhat theoretically supported) raises the gas mass by a factor of 5.}
\item{This size difference also manifests in the form of a star formation efficiency variation, with `sub clumps' of our sources differing in in efficiency by up to a factor of $\sim 5$. However, the global properties of each source are consistent with the canonical L\arcmin$_{\mathrm{CO}}$-L$_{\mathrm{FIR}}$ relation for similarly extreme (and sub-mm bright) systems.}
\item{These are the first PdBI observed `A' configuration observations of high-$z$ ULIRGs for CO(4$\rightarrow$3), achieving the smallest beam size yet for an intermediate-excitation transition at a redshift of $\sim$2. The modest exposure times required (6 antenna equivalent of $\sim$ 8 hours), the small loss of flux (a factor of 1.9 on average), and the detailed morphologies and kinematic information revealed suggest that this is a profitable approach for similar investigations in the future.}
\end{itemize}


\section*{Acknowledgments}
This study is based on observations made with the IRAM Plateau de Bure Interferometer. IRAM is supported by INSU/CNRS (France), MPG (Germany) and IGN (Spain). We acknowledge the use of \textsc{gildas} software (http://www.iram.fr/IRAMFR/GILDAS). This work also makes use of observations taken by the University of Manchester at Jodrell Bank Observatory on behalf of STFC, and the VLA of the National Radio Astronomy Observatory, a facility of the National Science Foundation operated under agreement by Associated Universities, Inc. We are grateful to the Great Observatories Origins Deep Survey (GOODS) team for use of their ACS data. We would like to thank the anonymous referee, whose comments and suggestions helped improve this work. We would also like to thank Mark Krumholz for his enlightening thoughts on the CO conversion factor. MSB and IRS acknowledge the financial support of STFC, and CMC thanks the Gates Cambridge Trust.

\label{lastpage}

\end{document}